\documentclass[preprint2]{aastex}

\shorttitle{S$^{4}$G Stellar Mass Maps}
\shortauthors{Querejeta et al.}

\usepackage{natbib}
\usepackage{graphicx}
\usepackage{tikz}
\usepackage{amsmath}
\usetikzlibrary{shapes,arrows}
\usepackage[top=2cm, bottom=3cm, left=2cm, right=1cm]{geometry}
\begin{document}

\title{The \textit{Spitzer} Survey of Stellar Structure in Galaxies (S$^4$G): Precise Stellar Mass Distributions from Automated Dust Correction at 3.6\,$\mu m$}

\author{
Miguel Querejeta\altaffilmark{1},
Sharon E. Meidt\altaffilmark{1},
Eva Schinnerer\altaffilmark{1},
Mauricio Cisternas\altaffilmark{2,3},
Juan~Carlos~Mu\~noz-Mateos\altaffilmark{4},
Kartik~Sheth\altaffilmark{5,6,7},
Johan~Knapen\altaffilmark{2,3},
Glenn~van~de~Ven\altaffilmark{1},
Mark~A.~Norris\altaffilmark{1},
Reynier~Peletier\altaffilmark{8},
Eija~Laurikainen\altaffilmark{9,10},
Heikki~Salo\altaffilmark{9},
Benne~W.~Holwerda\altaffilmark{11},
E.~Athanassoula\altaffilmark{13},
Albert~Bosma\altaffilmark{13},
Brent~Groves\altaffilmark{1},
Luis~C.~Ho\altaffilmark{12,20},
Dimitri~A.~Gadotti\altaffilmark{4},
Dennis~Zaritsky\altaffilmark{12},
Michael~Regan\altaffilmark{14},
Joannah~Hinz\altaffilmark{15},
Armando~Gil~de~Paz\altaffilmark{16},
Karin~Menendez-Delmestre\altaffilmark{17},
Mark~Seibert\altaffilmark{12},
Trisha~Mizusawa\altaffilmark{19},
Taehyun~Kim\altaffilmark{20},
Santiago~Erroz-Ferrer\altaffilmark{2,3},
Jarkko~Laine\altaffilmark{9},
S\'ebastien~Comer\'on\altaffilmark{9,10}}

\altaffiltext{1}{Max-Planck-Institut f\"ur Astronomie, K\"onigstuhl 17, D-69117 Heidelberg, Germany}
\altaffiltext{2}{Instituto de Astrof\'isica de Canarias, Spain}
\altaffiltext{3}{Departamento de Astrof\'isica, Universidad de La Laguna, Spain}
\altaffiltext{4}{European Southern Observatory, Chile}
\altaffiltext{5}{National Radio Astronomy Observatory, Charlottesville, USA}
\altaffiltext{6}{Spitzer Science Center, Pasadena, USA}
\altaffiltext{7}{California Institute of Technology, USA}
\altaffiltext{8}{Kapteyn Astronomical Institute, Groningen, The Netherlands}
\altaffiltext{9}{University of Oulu, Finland}
\altaffiltext{10}{Finnish Centre of Astronomy with ESO, Turku, Finland}
\altaffiltext{11}{Leiden Observatory, Leiden University, Leiden, The Netherlands}
\altaffiltext{12}{The Observatories of the Carnegie Institution for Science, Pasadena, USA}
\altaffiltext{13}{Laboratoire d'Astrophysique de Marseille (LAM), France}
\altaffiltext{14}{Space Telescope Science Institute (STScI), Baltimore, USA}
\altaffiltext{15}{MMT Observatory/University of
Arizona, Tucson, USA}
\altaffiltext{16}{Universidad Complutense de Madrid (UCM), Spain}
\altaffiltext{17}{Observat\'orio do Valongo, Brazil}
\altaffiltext{18}{Laboratoire d'Astrophysique de Marseille (LAM), France}
\altaffiltext{19}{Florida Institute of Technology, USA}
\altaffiltext{20}{Kavli Institute for Astronomy and Astrophysics, Peking University, Beijing, China}

\email{querejeta@mpia.de}

\begin{abstract}

The mid-infrared is an optimal window to trace stellar mass in nearby galaxies and the 3.6\,$\mu m$ IRAC band has been exploited to this effect, but such mass estimates can be biased by dust emission.
We present our pipeline to reveal the old stellar flux at 3.6\,$\mu m$ and obtain stellar mass maps for more than 1600 galaxies available from the \textit{Spitzer} Survey of Stellar Structure in Galaxies (S$^{4}$G).
This survey consists of images in two infrared bands (3.6 and 4.5\,$\mu m$), and we use the Independent Component Analysis (ICA) method presented in \citet{2012ApJ...744...17M} to separate
the dominant light from old stars and the dust emission that can significantly contribute to the observed 3.6\,$\mu m$ flux.
We exclude from our ICA analysis galaxies with low signal-to-noise ratio ($S/N < 10$) and those with original [3.6]-[4.5] colors compatible with an old stellar population, indicative of little dust emission (mostly early Hubble types, which can directly provide good mass maps). For the remaining 1251 galaxies to which ICA was successfully applied, we find that as much as 10-30\% of the total light at 3.6\,$\mu m$ typically originates from dust, and locally it can reach even higher values. 
This contamination fraction shows a correlation with specific star formation rates, confirming that the dust emission that we detect is related to star formation.
Additionally, we have used our large sample of mass estimates to calibrate a relationship of effective mass-to-light ratio ($M/L$) as a function of observed [3.6]-[4.5] color:  $\log(M/L)=-0.339 (\pm 0.057) \times ([3.6]-[4.5]) -0.336 (\pm 0.002)$.
Our final pipeline products have been made public through IRSA, providing the astronomical community with an unprecedentedly large set of stellar mass maps ready to use for scientific applications.

\end{abstract}


\section{Introduction}

Cosmological studies have revealed a close relation between stellar mass and star formation rates (SFRs) in galaxies \citep[e.g.,][]{2007ApJ...670..156D,2007ApJ...660L..43N,2011ApJ...735L..34G}, which implies that stellar mass controls to a great extent their growth and evolution. 
However, even more illuminating than the total mass of a galaxy is the actual spatial distribution of its baryonic mass. This snapshot of the
present-day gravitational potential constitutes a fossil record of the evolutionary history that led to its current state. Therefore, reliable maps of 
the stellar mass distribution provide a vital tool to probe the baryonic physics responsible for shaping galaxies to their present state.

Maps of the stellar mass distribution have played an important role in understanding many structural and evolutionary effects in galaxies that are thought to be linked to secular evolution \citep[see][for a review]{2004ARA&A..42..603K}. For example, stellar mass maps traced by near-IR imaging have been used to study torques exerted by the stellar structure \citep[e.g.,][]{1986ApJ...303...66Z,2010MNRAS.407..163F}. They are also critical to evaluate the role of bars \citep[e.g.,][]{1988ApJ...327L..61S, 1995ApJ...454..623K, 2000ApJ...529...93K, 2005ApJ...632..217S,2007ApJ...670L..97E,2010ApJ...715L..56S} and to confirm theoretical predictions for bar formation \citep[][]{1992MNRAS.259..328A,2013seg..book..305A}. Studies of the spiral structure \citep[e.g.,][]{1984ApJS...54..127E,1989ApJ...343..602E,1997AJ....114..965R} and inner stellar components in galaxies \citep[e.g.,][]{2002AJ....124...65E, 2006MNRAS.369..529F, 2011A&A...533A.104E} also benefit from the use of the true underlying stellar mass distribution.

Similarly, knowing how the stellar mass is organized is essential if we want to understand how the gravitational potential influences the gas distribution and dynamics. For instance, mass maps have been used to infer the gravitational torques acting on the gas to determine nuclear inflows or outflows \citep[see, for instance,][]{1995ApJ...454..623K,2005A&A...441.1011G,2005ApJ...630..837J,2008A&A...482..133H,2009ApJ...692.1623H}. In a recent study, knowledge of the gravitational torques has been applied to assess the role of dynamics and stability of giant molecular clouds in the nearby galaxy M51 \citep{2013ApJ...779...45M}. Therefore, stellar mass maps are an important tool when it comes to understanding the response of gas to the underlying gravitational potential.

While several strategies can be used to measure the total mass of a galaxy \citep[see recent review by][]{2014RvMP...86...47C}, unbiased stellar mass maps of galaxies are not trivial to obtain, and they  often involve large uncertainties. 
Optical images have been used along with prescriptions to calculate the
mass-to-light ratio \citep[M/L$ \equiv \Upsilon$, in the most sophisticated case obtained using two color images, see][]{2009MNRAS.400.1181Z}. Alternatively, kinematic information can be used to 
derive the total mass within a given radius \citep[e.g.][]{2006MNRAS.366.1126C}, which, in combination with a model for the dark matter halo, can provide
an independent measure of the baryonic mass distribution for a given galaxy. Newer strategies include fitting stellar population models to IFU spectral data-cubes, from which an estimate of surface mass density can be obtained \citep[such as those of CALIFA,][]{2013A&A...557A..86C}. 
In any case, the instrumental requirements or the methodological complexities of these strategies have so far prevented researchers from obtaining maps of the stellar mass distribution for large samples (of more than $\sim 100$ galaxies).

In this paper, we apply a novel technique to infrared images, first introduced in \citet{2012ApJ...744...17M}, which allows us to automatically produce high-quality mass maps for
a large fraction of the galaxies in the \textit{Spitzer} Survey of Stellar Structure in Galaxies \citep[S$^{4}$G,][]{2010PASP..122.1397S}. S$^{4}$G provides deep imaging for 2352 nearby
galaxies ($D<40$~Mpc) at the wavelengths of 3.6 and 4.5\,$\mu m$, probing stellar surface densities down to $\sim 1M_{\odot}/pc^2$. The near- to mid-infrared regime provides a very good window to trace stellar mass, as the light emitted at these wavelengths is dominated by K- and M-type giant stars, tracing the old stellar populations that dominate the baryonic mass budget of nearby galaxies \citep[e.g.][]{1993ApJ...418..123R}.
As extinction is a strong function of wavelength, any stellar light emitted in the mid-infrared is significantly less affected than 
in the optical or at shorter wavelengths. Consequently, these data allow us to probe the mass distribution very well even in moderately inclined galaxies. While extinction is of little concern, emission from dust can significantly contribute to the flux detected in the 3.6 and 4.5\,$\mu m$ filters, in particular from the 3.3\,$\mu m$ PAH feature and hot dust arising from massive star-forming regions or around active nuclei \citep{2012ApJ...744...17M}.

In order to effectively separate the old stellar light from this dust emission which has a very different spectral energy distribution from the stars, we use a method based on Independent Component Analysis (ICA). Our method typically identifies the main dust contribution at 3.6 and 4.5\,$\mu m$ \citep[diffuse dust;][]{2012ApJ...744...17M}
and also allows one to remove the localized flux arising from circumstellar dust related to the late phases of red stars with lower $M/L$ \citep{2012ApJ...748L..30M}.
After removing this dust emission, we are left with a smooth distribution of essentially old stars (age~$\tau \sim$~2-12\,Gyr). According to \citet{2014ApJ...788..144M} and \citet{2014ApJ...797...55N}, the age and metallicity dependence of the $M/L$ at 3.6\,$\mu m$ for old stars is so modest that even a single M/L$ \equiv \Upsilon_{3.6}=0.6$ provides a conversion into mass accurate within $\sim 0.1$\,dex.

We have constructed a pipeline (S$^{4}$G Pipeline 5) to automatically remove dust emission from the 3.6\,$\mu m$ images in the large S$^4$G survey of galaxies. Application of an appropriate $M/L$ \citep[see][]{2014ApJ...788..144M} to these cleaned maps results in stellar mass maps. All the maps of the old stellar light, along with those of the identified dust emission, will be made publicly available through the NASA/IPAC Infrared Science Archive (IRSA).\footnote{Webpage: irsa.ipac.caltech.edu}

In this paper we explain the details of our pipeline and analyze some properties of the two components identified by ICA for the wide range of galaxies present in S$^4$G.
After presenting the data in Sect.~\ref{data}, the ICA method to remove dust emission is physically motivated and briefly explained in Section~3, followed by our description of the detailed pipeline in Section~4. Our iterative approach of ICA is presented here, along with a brief justification of the different steps involved in the process and an account of the uncertainties.
In Section~5 we describe the conversion to mass maps and the public product release. Section~6 is an account of the final sample and general properties of solutions, whereas results are presented Section~7. We close the paper with a summary and some conclusions in Section~8.

\section{Observations and Data Processing}
\label{data}

S$^{4}$G \citep{2010PASP..122.1397S} has imaged 2352 galaxies using the Infrared Array Camera
\citep[IRAC,][]{2004ApJS..154...10F} at 3.6 and 4.5\,$\mu m$, the bands which are still available within the 
post-cryogenic mission of the \textit{Spitzer Space Telescope}. The selection of galaxies corresponds to a volume ($d < 40$~Mpc), 
magnitude ($m_{B,corr} < 15.5$~mag) and size limit ($D_{25} > 1$'), and provides deep images reaching $\mu_{3.6\mu m}(AB)(1\sigma)\sim 27~$mag$~$arcsec$^{-2}$ (about $1M_{\odot}/pc^2$).

All the images have been processed using the S$^{4}$G pipeline, which consists of five steps. The first four steps are summarized in \citet{2010PASP..122.1397S},
and are described in detail in two companion papers \citep{2015ApJS..219....3M,2015ApJS..219....4S}. The process that we present here corrects for emission from dusty sources in the IRAC bands, and it is the fifth stage of the S$^{4}$G pipeline system:

\begin{itemize}
\item \textbf{Pipeline 1} transforms the raw data into science-ready {\tt FITS} images by mosaicking and matching the background levels, providing a resulting 
pixel scale of 0.75 arcsec/pixel \citep{2015ApJS..219....3M}. The point-spread function (PSF) has a typical FWHM of 1.7'' and 1.6'' at 3.6 and 4.5\,$\mu m$, respectively.

\item \textbf{Pipeline 2} generates masks, first based on {\tt SExtractor} identifications, and then checked and modified by eye \citep{2015ApJS..219....3M}.

\item \textbf{Pipeline 3} takes care of measuring sky levels, determining the center and obtaining isophotal values \citep[intensity, 
surface brightness, ellipticity, position angle;][]{2015ApJS..219....3M}.

\item \textbf{Pipeline 4}, performs {\tt GALFIT} photometric structural decompositions of the sample \citep{2015ApJS..219....4S}.

\item \textbf{Pipeline 5}, presented here, relies on previous pipeline steps and is the application of the ICA method to separate the light from old stars and dust.

\end{itemize}

Additionally, at some points in this paper, and especially for the analysis of results (Sect.~\ref{sec:results}), we will make use of a set of ancillary information. For distances, whenever available, we rely on the mean redshift-independent distance provided by the NASA Extragalactic Database (NED); otherwise, they are based on the observed radial velocity, also from NED
\citep[compiled by ][]{2015ApJS..219....3M}.
We utilize the morphological classification at 3.6\,$\mu m$ from \citet{2010ApJS..190..147B} and \citet{2015ApJS..217...32B}.
SFRs are derived from IRAS photometry at 60\,$\mu m$ and 100\,$\mu m$ obtained from NED following \citet{2000A&A...354..836L}.

\section{The ICA Technique}
\subsection{Expected Sources at 3.6 and 4.5\,$\mu m$ and Their Colors}
\label{Expected_sources}
For nearby galaxies, the light detected in the IRAC 3.6\,$\mu m$ filter arises mainly from two components: the photospheres of (old) stars and from dust emission. Old stars dominate the stellar flux, as their atmospheres are cold and their blackbody curves peak close to 3.6\,$\mu m$; additionally, due to CO absorption at 4.5\,$\mu m$, old K and M giants exhibit blue [3.6]-[4.5] colors \citep{2004ApJS..154..235P,2004ApJS..154..222W}. Younger (hotter) stars are not expected to contribute significantly to the observed stellar emission. However, due to their strong UV fluxes, these younger stars can heat their surrounding dust, which, in turn, re-radiates at longer wavelengths and can also account for a significant fraction of the light at 3.6\,$\mu m$ \citep{2012ApJ...748L..30M}. Models show that the dust emission arises from the PAH bands (specifically, the 3.3\,$\mu m$ feature) and the continuum radiation from very small grains \citep[e.g.][]{2001ApJ...551..807D}, and this (hot) dust emission becomes more prominent near the sites of star formation, where the presence of young stars leads to stronger radiation fields.

\begin{figure}[ht!]
\begin{center}
 \includegraphics[width=0.48\textwidth]{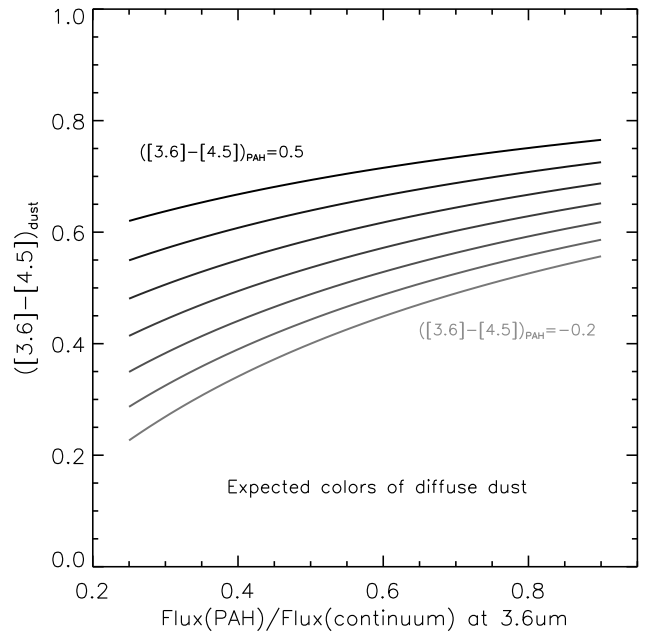}
 \caption{Observed [3.6]-[4.5] color as a function of the relative contribution to flux from diffuse dust continuum and PAH. The different curves correspond to different intrinsic PAH colors ($[3.6]-[4.5]\vert_{\mathrm{PAH}}=-0.2$ to 0.5 in steps of 0.1). For reasonable fractions of such a mixture, we find that the resulting dust color is constrained within the range 0.2--0.7 (see text for details).}
\label{fig:dustcolors}
\end{center}
\end{figure}

According to the measurements of \citet{2006A&A...453..969F}, PAHs should have colors [3.6]-[4.5]$\sim$0.3, including both the 3.3\,$\mu m$ PAH feature and the underlying PAH continuum detected, e.g., in Milky Way reflection nebulae.  In the absence of the PAH continuum, the [3.6]-[4.5] color is bluer (as low as [3.6]-[4.5]$\sim$-0.1), although this scenario will likely be uncommon here, given unavoidable mixing at our resolution.  For the continuum dust component, we expect [3.6]-[4.5]$\sim$1.0, adopting the power-law approximation $f_\nu \propto \nu ^ {- \alpha}$ to the Wien side of the dust spectral energy distribution (SED) with $\alpha = 2$, which \citet{2003MNRAS.338..733B} suggest better accounts for the observed dust SED shape than a modified blackbody in the near-IR. As shown in Fig.~\ref{fig:dustcolors}, the diffuse mixture of PAH and continuum dust emission will exhibit resulting colors in the range 0.2$<$[3.6]-[4.5]$<$0.7.

The relative flux contributions of the PAH and continuum at 3.6\,$\mu m$ can be estimated from the ratio of the 6.2\,$\mu m$ PAH feature to the 6.2\,$\mu m$ continuum measured by \citet{2002A&A...390.1089P}. For normal spiral galaxies, this ratio is between 0.72-1.16 (on average $\sim$1) and can be as low as 0.1 in QSOs and the nuclei of Seyfert galaxies.  The ratio of the 3.3\,$\mu m$ to 6.2\,$\mu m$ PAH features measured in HII regions, star-forming regions and in planetary nebulae falls in the range 0.15-0.25 \citep{2011A&A...531A.137H}.  With the Blain et al. power-law dust continuum, this implies that the ratio of the 3.3\,$\mu m$ PAH emission to the underlying continuum is $\sim$0.4-0.9, or as low as 0.1 in galaxy nuclei or other regions dominated by hot dust.  

Fig.~\ref{fig:dustcolors} shows the resulting colors for the diffuse dust at 3.6\,$\mu m$ adopting this range of PAH flux fractions for a wide range of possible PAH colors and assuming a fixed continuum color. 
As found by \citet{2012ApJ...744...17M}, the primary non-stellar emission detectable with ICA is in the form of this `diffuse dust' component, i.e. the mixture of PAH and the dust heated by the ambient interstellar radiation field, away from star-forming regions.  As considered later in Sect.~\ref{ICA2} and in Appendix~B, some galaxies also exhibit additional emission from hot dust, isolated in star-forming HII regions.  This dominant component in HII regions, which can be thought of as the far end of the spectrum exhibited by the `diffuse dust', should have [3.6]-[4.5] colors closer to 1.

In the implementation of our S$^4$G pipeline we make use of the fact that regions containing hot dust (and negligible PAH) have colors that are distinguishable from the nominal diffuse dust component.  We also make use of the fact that the colors of the dust and the old stellar population are very different.  
The colors for old stellar populations lie in the range $-0.2<[3.6]-[4.5]\vert_{\mathrm{stars}}<0$, according to the observed colors of giant stars \citep{2014ApJ...788..144M}. This is consistent with the observed colors of early-type galaxies, in the absence of significant dust emission \citep[][and \citet{2014ApJ...797...55N}]{2012MNRAS.419.2031P}. The non-stellar $[3.6]-[4.5]$ color, on the other hand, always appears positive in these bands ($[3.6]-[4.5]\vert_{\mathrm{dust}}>0$), although it can span a wider range, as seen in Fig.~\ref{fig:dustcolors}. 

\begin{table}[t!]
\begin{center}
\caption[h!]{Main Sources at 3.6\,$\mu m$ and Their [3.6]-[4.5] Colors.}
\begin{tabular}{cc}
\tableline\tableline
 Source & Typical $[3.6]-[4.5]$ Range\\   
\tableline
 Old stars \tablenotemark{\it a} & $-0.2 \mathrm{~...~} 0$   \\
  Diffuse dust  & $\sim 0.2 \mathrm{~...~} 0.7$ \\
 PAH emission \tablenotemark{\it b} & $\sim 0.3$     \\
 Dust in HII regions \tablenotemark{\it c} & $\sim 1.0$ \\
 \tableline
\end{tabular}
\label{table:colsources}
\end{center}
\tablecomments{Main sources identified with ICA, and typical values \\ of their colors.
}
\tablenotetext{a}{\citet{2004ApJS..154..222W}, \citet{2004ApJS..154..235P}, \citet{2012MNRAS.419.2031P},  \\  \citet{2014ApJ...788..144M}, \citet{2014ApJ...797...55N}}
\tablenotetext{b}{Dominated by PAH emission. Approximate estimation based on \\ \citet{2006A&A...453..969F}.}
\tablenotetext{c}{Representative number based on \citet{2003MNRAS.338..733B}; it can either \\ correspond to hot dust, e.g., in HII regions, or, in most extreme \\ cases, to hot dust heated to large temperatures near an AGN.}
\end{table}

\begin{figure}[h!]
\begin{center}
 \includegraphics[width=0.48\textwidth]{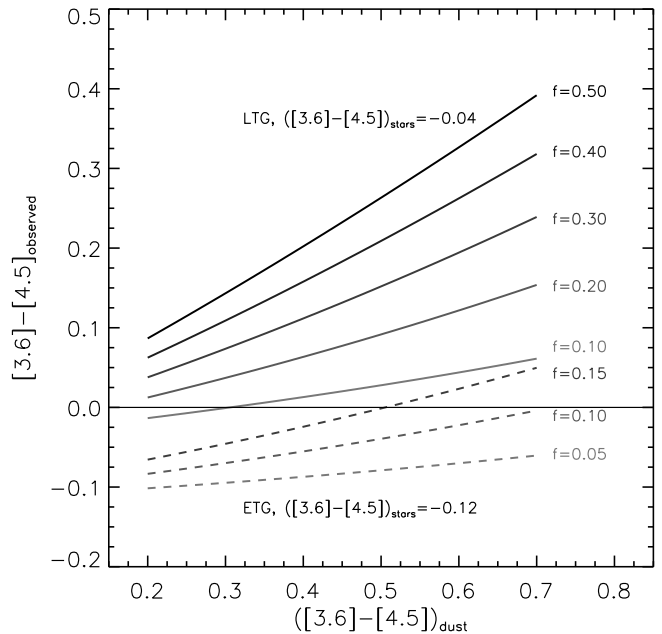}
 \caption{Observed [3.6]-[4.5] colors that arise from mixing stars with dust emission. Stars have colors $-0.2<[3.6]-[4.5]\vert_{\mathrm{stars}}<0$, and dust emission typically covers the range $0.2 \lesssim [3.6]-[4.5]\vert_{\mathrm{dust}} \lesssim 0.7$.
Here, we show the resulting color for realistic combinations of dust colors and fractions, for two fixed stellar colors. One of them is representative of late-type galaxies (LTG, $[3.6]-[4.5]\vert_{\mathrm{stars}}=-0.04$), and the other one is standard of early-type galaxies (ETG, $[3.6]-[4.5]\vert_{\mathrm{stars}}=-0.12$). The fractions of dust emission at 3.6\,$\mu m$ ($f$) are also limited to the maximum expected fractions in early- and late-type galaxies, as estimated in Appendix~A. Virtually all possible combinations lead to global negative colors in the case of early-type galaxies, whereas essentially all realizations produce positive global colors for late-type galaxies, in which the contribution from dust can be more significant.}
\label{fig:origcolors}
\end{center}
\end{figure}

Table~\ref{table:colsources} summarizes the colors expected for the different sources of emission at 3.6\,$\mu m$.  These sources combine together to produce the observed [3.6]-[4.5] colors plotted in Fig.~\ref{fig:origcolors}.  There we show representative colors for young, relatively metal-poor, late-type galaxies and their old, metal-rich early-type counterparts (with bluer [3.6]-[4.5] colors)\footnote{As a result of CO absorption in the 4.5\,$\mu m$ band, the metallicity dependence of the [3.6]-[4.5] color identified by \citet{2014ApJ...788..144M} is the reverse of that exhibited by optical or optical-NIR colors; the [3.6]-[4.5] color becomes more blue with increasing metallicity \citep[see also][]{2014ApJ...797...55N}.}, adopting the dust contamination fractions estimated in Appendix~A.  The latter consistently exhibit [3.6]-[4.5]$<$0, whereas the addition of even a little bit of dust emission will move the [3.6]-[4.5] colors of late-type galaxies above zero.

\subsection{Independent Component Analysis}

Here, we introduce the ICA method and the nomenclature that we will use. The method has already been described and validated by \citet{2012ApJ...744...17M}, so we will only provide a short summary here. 

Similar to Principal Component Analysis (PCA), Independent Component Analysis (ICA) is a means of \textit{blind source separation} which we use to extract measurements of the flux and 
the wavelength-dependent scaling of individual components from linear combinations of the input data. But in contrast to PCA, ICA maximizes the statistical independence of 
the sources rather than requiring that the sources are orthogonal (with zero covariance).  We use the {\tt fastICA} realization of the method developed by \citet{1999ISPL....6..145H} 
and \citet{Hyvarinen_Oja}, which achieves statistically independent solutions by maximizing the non-Gaussianity of the source distributions.  

In practice, given $N$ input images (e.g., at $N$ different wavelengths), ICA will identify at most $N$ underlying distinct sources that contribute to the flux in each of the $N$ images, under the assumption that each image is a linear combination of the sources.  Because 
S$^{4}$G provides images of each galaxy in two channels, ICA can identify two distinct components through solution of the following equation: 

\begin{equation}
 \mathbf{x} = \mathbf{A} \cdot \mathbf{s},
\end{equation}

\noindent
where \textbf{x} is the $2 \times P$ measurement set, with as many columns as pixels $P$ in the analysis region, \textbf{s} is the $2 \times P$ source solution set, and 
$A_{i,j}$ is an invertible $2 \times 2$ matrix of `mixing coefficients' determined simultaneously with $\textbf{s}$.  

As first considered by \citet{2012ApJ...744...17M}, ICA provides a way to distinguish between the old stellar 
population that dominates the light in the IRAC bands and additional emission present to varying degrees (depending on the nature of the  source), without {\it a priori} knowledge of the relative proportions of the sources or their colors between 3.6 and 4.5\,$\mu m$.
In six prototypical star-forming disk galaxies, \citet{2012ApJ...744...17M} showed that ICA can identify and remove the combined PAH and continuum dust emission
 (tracing star formation, as observed at longer wavelengths) and localized emission from dusty asymptotic giant branch (AGB) and red supergiant (RSG) stars.  The ICA correction leaves a cleaned, smooth map of the old stellar light consistent with expectations for an old, dust-free stellar population.  
 
 Our pipeline implementation of ICA includes an estimate of the uncertainties on both components by running the ICA sequence 48 times, based on 48 perturbations to the mixing matrix. As an initial guess for the mixing coefficients we adopt the expected color range of stars and dust ($-0.2 < [3.6] - [4.5]\vert_{\mathrm{stars}} < 0$ and $0 < [3.6] - [4.5]\vert_{\mathrm{dust}} < 1.5$),
but we find that ICA solutions quickly depart from the initial seeds and converge to the final solution. The perturbations to the mixing matrix correspond to small steps in dust and stellar color in a count-controlled nested loop. The primary loop implements a change in the stellar color by 0.04~mag, during which the secondary iterates the dust color in steps of 0.3 mag (at fixed stellar color).

\subsection{Notation}
For each of the two original input images we obtain two source images, which we refer to as $s1$ and $s2$, providing a total of four images (which we could in principle name $s1_{3.6\mu m}$, $s2_{3.6\mu m}$ and $s1_{4.5\mu m}$, $s2_{4.5\mu m}$).  At either wavelength it will always hold that $(s1 + s2) = \{\mathrm{original\;image} \} $.  
However, since the two source images at 4.5\,$\mu m$ are scaled replicas of the two source images at 3.6\,$\mu m$, with scalings set by the colors of the solutions ($[3.6] - [4.5]\vert_{s1}$ and $[3.6] - [4.5]\vert_{s2}$), throughout the
paper we will, by default, refer to the $3.6 \mu m$ solutions as $s1$ and $s2$.

\begin{figure*}[ht!]
\begin{center}
 \includegraphics[trim = 0mm 135mm 0mm 0mm, clip, width=0.7\textwidth]{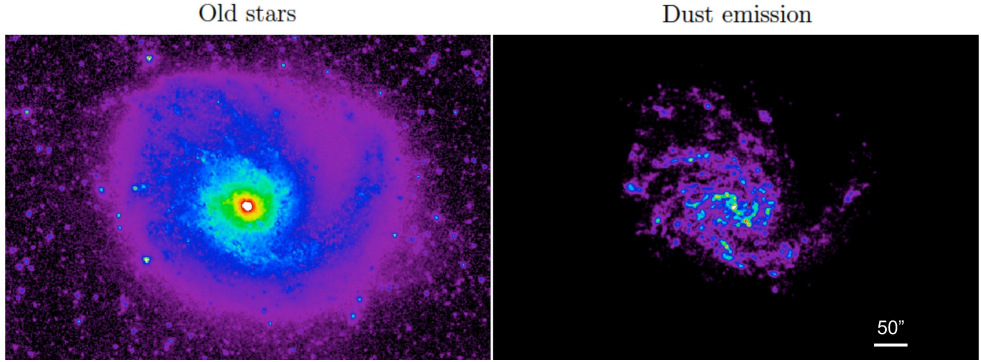}
 \caption{NGC\,4254 as a prototypical example of the ICA decomposition of the
 3.6\,$\mu m$ image into old stars ($s1$, left) and dust emission ($s2$, right). This is the standard case for star-forming disk galaxies, with a dust color of $[3.6]-[4.5]\vert_{s2}=0.17$ and finding a color of the old stellar component of $[3.6]-[4.5]\vert_{s1}=-0.02$, in agreement with the values expected for K and M giants. Here, and throughout the paper, $s1$ maps are shown in square root intensity scale, with $I^{max}_{3.6}=10$\,MJy/sr, whereas $s2$ maps are displayed in linear scale, with $I^{max}_{3.6}=5$\,MJy/sr (in both cases $I^{min}_{3.6}=0$).}
\label{fig:n628}
\end{center}
\end{figure*}

With this notation, 
ICA typically separates the emission arising from old stars (which we identify with $s1$) from dust ($s2$), as is illustrated by Fig.~\ref{fig:n628}. The secondary emission that arises from dust can have a number of physical origins, covering different color ranges (see Table~\ref{table:colsources}), but is always redder than old stars in these bands, which allows ICA to perform the separation into two distinct components.

\section{S$^{4}$G Pipeline 5}
\label{applyica}

In this Section we summarize our approach for applying the ICA technique to the S$^{4}$G sample, exploiting the powerful resources available from S$^{4}$G (images at 3.6 and 4.5\,$\mu m$ and pipeline products). We introduce and describe our iterative implementation of ICA, which is designed to ignore all pixels that contain emission different from the dominant two (e.g., background galaxies, foreground stars), thus allowing ICA to produce better results.

Given the size and diversity of the sample, it is not surprising that there are some galaxies for which we cannot expect to apply ICA under optimal conditions.
In the same way that a low signal-to-noise ratio (S/N) can degrade the quality of color information that can be extracted from 2D images, low S/N provides inadequate leverage on the spectral shapes of the two components between 3.6 and 4.5\,$\mu m$ which is necessary for ICA to obtain a robust solution.
Our testing of the method confirms that ICA cannot correctly separate images into two components when the average signal-to-noise is low.\footnote{Our analysis of the set of galaxies with S/N below 10 confirms that S/N is a major determinant in the quality of solutions; we find that 67\% of these cases have solutions with an $s1$ component with colors outside the expected range for old stars. We therefore leave out those galaxies with the poorest data quality, by definition.}
Therefore, we have adopted a conservative approach and apply our pipeline only to those galaxies with average signal-to-noise above $S/N>10$.

\begin{figure}[h!]
\begin{center}
 \includegraphics[trim = 40mm 70mm 40mm 40mm, clip, width=0.48\textwidth]{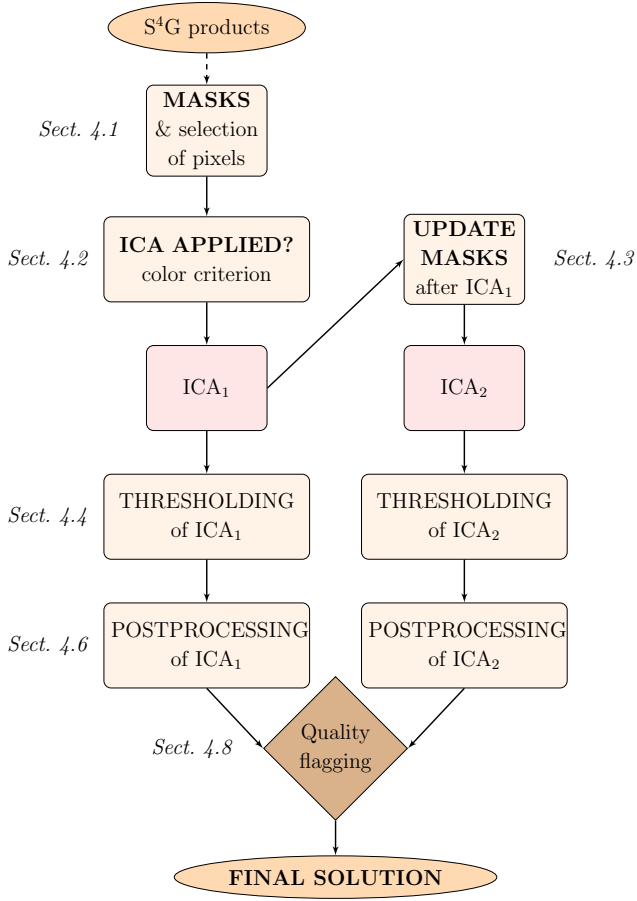}
 \caption{Flow diagram showing the different steps involved in S$^4$G Pipeline~5. The blocks shaded in light orange are executed in {\tt IDL}, whereas the parts in light red correspond to {\tt C++}.}
\label{fig:flowdiagr}
\end{center}
\end{figure}

Obtaining dust-free flux maps in an automatic fashion with ICA involves a number of steps (masking, second ICA iteration, thresholding, postprocessing) that will be introduced and justified in the present section. 
For clarity, the steps involved in the pipeline are presented in a flow chart (Fig.~\ref{fig:flowdiagr}), with a reference in {\it italics} to the specific section where they are discussed.

\subsection{Defining the ICA solution area}
\label{subsec:solarea}

Because our implementation of ICA seeks to maximize the non-Gaussianity of the source distributions \citep{Hyvarinen_Oja},
ICA is sensitive to extreme outliers in color, even if these cover only a small region. This means that a few pixels with very different color from 
the dominant color can bias the whole solution (not only the area they cover), yielding an unrepresentative separation of the original image into two components. Consequently, properly masking any foreground stars or background galaxies is particularly important.

We have developed a specific masking strategy that builds on masks already available from S$^{4}$G Pipeline~2, but includes further corrections to ensure that bright regions belonging to the galaxy are not masked in a first attempt.
Pipeline~2 masks were developed primarily for galaxy photometric decompositions (Pipeline~4), and that is the reason why this subtle yet important modification is necessary to make the masks applicable for our purposes here.

Specifically, we remove from the Pipeline~2 masks any contiguous regions that are small (maximum area in pixels of $8000 \mathrm{Mpc^2}/d^2$) and have colors $-0.3<[3.6]-[4.5]\vert_{\mathrm{orig}}<0.75$, as they tend to
be bright regions intrinsic to the target galaxy (e.g., HII regions). The distance-dependent cut in size corresponds to a
radius of $\sim150$~pc, covering well even the largest HII regions \citep[see][]{2011ApJ...729...78W}, whereas the color criterion
matches roughly the range spanned by emission from old stars and dust, leaving out saturated field stars and observational artifacts, which usually exhibit colors outside this range.

Therefore, the set of pixels to which we apply ICA (the ICA \textit{solution area}) is defined to contain all emission out to the edge of the galaxy and avoid external sources such as field stars and background galaxies. The area is centered on the photometric center defined in Pipeline~3 and extends out to the 25.5 mag~arcsec$^{-2}$ isophote \citep{2015ApJS..219....3M}, excluding the masked regions. At this early stage, we also correct for PSF effects by convolving both images with the IRAC 4.5\,$\mu m$ to 3.6\,$\mu m$ PSF kernel from \citet{2011PASP..123.1218A}.

\subsection{Deciding Whether ICA Should be Applied}
\label{subsec:icaneed}

The expected color for an old stellar population of ages $\tau \sim$~2-12~Gyr is $-0.2<[3.6]-[4.5]<0$ (see references in Table~\ref{table:colsources}). Some of the galaxies in S$^4$G have original [3.6]-[4.5] global colors in that range, which implies that they are already compatible with an old stellar population. As shown in Fig~\ref{fig:origcolors}, an originally blue color is indicative of little to no dust emission; we therefore do not apply ICA to these galaxies.
  We have found that these cases can be best identified by calculating the \textit{weighted mean} of the original color for the pixels in our ICA solution area; the weights are chosen as the inverse of the variances ($w_i = 1 / \sigma_i^2$, where $\sigma_i$ is the original color error, as described in Sect.~\ref{subsect:uncertainties}).  By adopting the weighted mean we avoid the bias of low signal-to-noise regions, from which color information is less reliable. A total of 376 galaxies in S$^4$G (16\% of the sample) have colors originally compatible with old stars, and are therefore excluded from the further ICA analysis. 

Inspection of the 2D color maps confirms very little contamination from dust in galaxies where the original negative color is consistent with that of old stars, although dust may not be entirely absent.  Still, we prefer not to run ICA in these cases, since our testing suggests that more uncertainty in the old stellar light map can be introduced under these circumstances than if the non-stellar emission is simply retained.  In particular, we have found that when emission from dust becomes negligible, the separation starts to be dominated by spatial fluctuations in the original color due to noise.  As shown in Appendix~A, a low fractional dust contamination leads to the largest errors on the stellar flux obtained from the ICA separation.

\subsection{Iterative ICA: Reducing the Number of Sources of Emission}
\label{ICA2}

As described in Sect~\ref{Expected_sources}, the primary non-stellar emission detectable with ICA is in the form of a `diffuse dust' component, i.e. a mixture of PAH and dust continuum. Some galaxies also exhibit secondary non-stellar emission from hot dust isolated in star-forming HII regions or near an active nucleus.  Even in these cases, it is most common that the diffuse dust, which is spatially more extensive than the dust in HII regions, dominates the ICA solution.  But as the hot dust regions are assigned a color that is unrepresentative (and, in particular, less red than their true color), the flux in this secondary dust component can be overestimated (see Sect.~\ref{subsec:solpostpro} below).  

In other cases (e.g. a high number of HII regions, or very pervasive star formation), ICA will favor the hot dust.  While this may provide the best description of the non-stellar emission in some galaxies (with genuinely more hot than diffuse dust), often the emission from the diffuse dust component is still present, but underestimated by ICA (since it is assigned a color redder than its intrinsic value).  
As we are more interested in isolating dust emission in the disk (tracing star formation, heating from hot evolved stars, etc.), we prefer to completely avoid these very red, hot dust-dominated regions by performing a second iteration of ICA with the corresponding areas masked.  

\begin{figure*}[ht!]
\begin{center}
 \includegraphics[trim = 0mm 45mm 0mm 0mm, clip, width=0.7\textwidth]{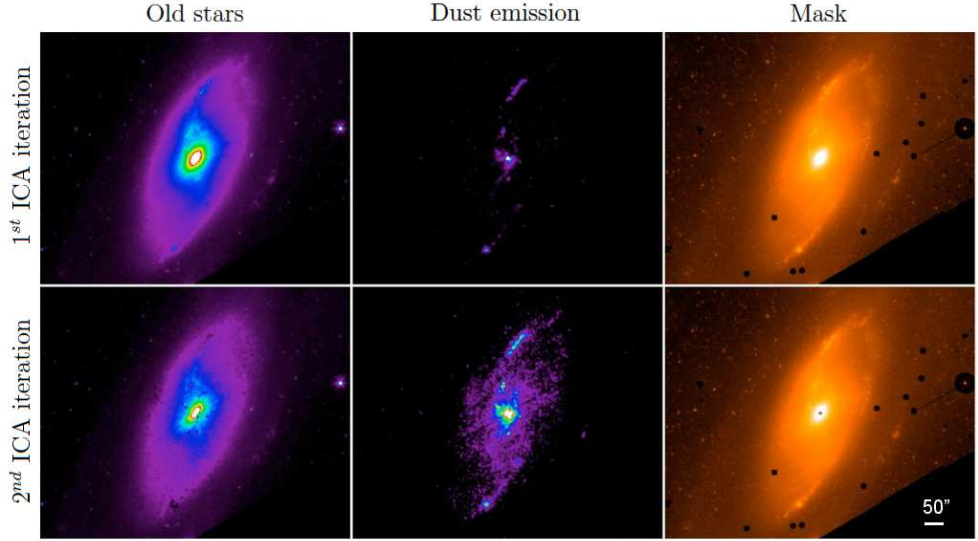}
 \caption{An example of how a second iteration of ICA with extreme color and flux outliers masked can reveal different components. Here, the first (top) and second (bottom) iterations are presented for NGC\,4258, a LINER galaxy. The left map corresponds to the old stellar flux ($s1$), the central column represents the dust map ($s2$), and the right column shows the mask used in each case (regions attenuated by 5\,mag against the original $3.6 \mu m$ image). For the first iteration, ICA finds a dust color of $[3.6]-[4.5]\vert^{\mathrm{ICA}_1}_{s2}=0.697$ and a stellar color of $[3.6]-[4.5]\vert^{\mathrm{ICA}_1}_{s1}=-0.028$. The second iteration effectively makes the dust color less red ($[3.6]-[4.5]\vert^{\mathrm{ICA}_2}_{s2}=0.138$), and the final stellar color is then $[3.6]-[4.5]\vert^{\mathrm{ICA}_2}_{s1}=-0.063$.}
\label{fig:sombrero}
\end{center}
\end{figure*}

This same iterative process is useful in general for reducing the number of sources of emission present in the \textit{solution area}. 
Extremely bright nuclear point-like sources have a very similar effect on ICA solutions as the bright field stars and background galaxies described in Sect.~\ref{subsec:solarea}.  In some cases, they initially completely dominate the secondary component identified with ICA.  
Ignoring the central source helps ICA identify other types of secondary sources of emission, as can be seen in Fig.~\ref{fig:sombrero}.

The relative contributions of the additional sources determines the degree to which ICA identifies a realistic secondary source.  
When, for example, the second component is dominated by the diffuse dust emission, and hot dust
appears in only small localized HII regions, ICA successfully describes the dominant source \citep[in this case diffuse dust; see][and Appendix~B]{2012ApJ...744...17M}.  This is close to the optimum decomposition and is characteristic of most star-forming disk galaxies. 
However, when the contributions of the two sources become comparable (e.g., when they cover similarly sized areas in the disk), ICA finds a compromise between the two. In some cases, this compromise is an acceptable outcome,
but it is often possible to obtain a noticeably improved solution for one of the components by running ICA again, now with pixels containing the other source masked.  
This is confirmed by the tests described in Appendix~A, and, for the interested reader, the details of this empirically optimized strategy are described below (Sect.~\ref{subsec:iterativeicadetails}).
 
In our automatic pipeline, the second ICA iteration is determined to be effective when the second ICA $s2$ is bluer than the first ICA $s2$ (i.e. the reddest sources have been removed from the dust map). This is the case for 66\% of the objects, in which $[3.6]-[4.5]\vert_{s2}$ is effectively reduced after the second iteration and ICA$_2$ is chosen. The improvement in the solution for these cases is also associated with a decrease in the color uncertainties (in 79\% of those objects $[3.6]-[4.5]\vert^{\mathrm{ICA_2}}_{s2}$ has a smaller uncertainty than $[3.6]-[4.5]\vert^{\mathrm{ICA_1}}_{s2}$). Even in these cases, the fractional change to the total old stellar flux (and, thus, total mass) is not dramatic, typically on the order of $\sim 5$\%, but of course this can locally become more significant.

We have confirmed that running a third iteration of ICA would introduce a change which is negligible, smaller than the uncertainty in 98\% of the cases. Therefore, for simplicity, we have only implemented two iterations in the final pipeline.

\subsubsection{Technical details on second iteration, ICA$_2$}
\label{subsec:iterativeicadetails}
 
 In practice, all the pixels with original colors redder than
the global color of the dust component ($[3.6]-[4.5]\vert_{s2}$) are selected. This is, by definition, equivalent to selecting the pixels in which oversubtractions in the stellar map have occurred (i.e.
negative values in the stellar map after the first iteration of ICA). To prevent masking any spurious isolated red pixels (i.e. those that arise from local fluctuations at low signal-to-noise), the regions identified as oversubtractions are radially dilated by 1 pixel.  We then check if, after the dilation, contiguous pixels still correspond to a region with original integrated color greater than $[3.6]-[4.5]\vert_{s2}$. If this is the case, the given region will be masked for the second iteration. 

Additionally, we analyze the distribution of fluxes in the map of dust and mask 
all the regions that are above $5\sigma$ ($\sigma$ meaning now the standard deviation in the distribution of the identified dust fluxes) if a given contiguous region has an
integrated color that exceeds either $[3.6]-[4.5]\vert_{s2}$ or an empirically set limit of $[3.6] - [4.5]\vert_{\mathrm{orig}}=0.1$.

Finally, to account for the reddest nuclear dust emission, often related to active galactic nucleus (AGN) activity, we perform one more modification on the masks of the galaxies in which the first ICA solution had an $s1$ color redder than old stars ($[3.6]-[4.5]\vert_{s1}>0$). Such a red $s1$ color indicates that the spatially dominant diffuse dust emission has not yet been removed with ICA, as an even redder nucleus is identified as virtually the only component by ICA in the $s2$ map.  
Therefore, in the second iteration, we mask the contiguous region in the $s2$ map which has the highest average flux, provided that it covers an area smaller than 200 px. This is about the maximum area we can expect to be covered by the PSF \citep{2011PASP..123.1218A} due to a point source; if the area is larger, we instead mask the circular region with a diameter of 10 px that maximises the $s2$ flux contained.

We confirm that most of the galaxies with $[3.6]-[4.5]\vert_{s1}^{\mathrm{ICA}_1}>0$ can be attributed to AGN activity \citep[55\% of these are classified as AGN in Simbad, whereas the global fraction of such AGN-classified galaxies in S$^4$G is only 8\%; see][]{2010A&A...518A..10V}. A substantial portion of the remaining 45\% has, conversely, been identified to have nuclear star formation (so-called HII nuclei). In any case, masking that central region allows ICA to identify the more extended dust emission, and in 70\% of the objects, this second iteration makes the $s1$ color match the expected values for an old stellar population (shifting from $[3.6]-[4.5]\vert_{s1}^{\mathrm{ICA}_1}>0$ to $[3.6]-[4.5]\vert_{s1}^{\mathrm{ICA}_2}<0$).

\subsection{Determining significant secondary emission: thresholding}
\label{subsec:sigemission}
ICA provides a solution for each pixel in the \textit{solution area} using information from all other pixels in the area.  But not all pixels contain a genuine secondary source; 
consequently, some low-level noise is systematically introduced 
throughout the analysis area. 
These low-level pixels can either be positive or negative, since ICA does not impose non-negative solutions, and they are clearly artificial, as they fluctuate nearly uniformly around zero. To prevent arbitrary removal (or addition) of flux from the stellar map, we conservatively impose a minimum flux of $s2$ above which the emission is arguably genuine. 
We base our threshold on the map of propagated uncertainties, which we are able to compute following the recipe described and justified below.

\subsubsection{Technical details on thresholding}
\label{subsec:sigemissiondetails}

\begin{figure}[h!]
\begin{center}
 \includegraphics[trim = 10mm 0mm 10mm 0mm, clip, width=0.54\textwidth]{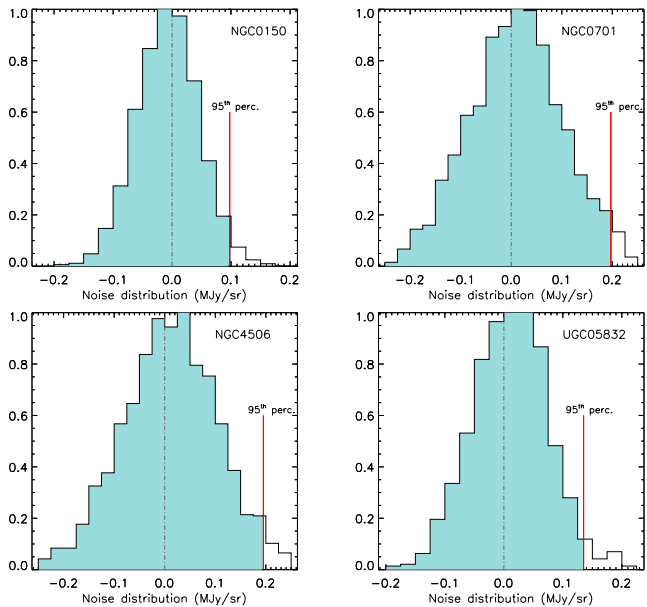}
 \caption{Histograms showing the flux distribution of $s2$ on areas where virtually no dust emission is expected (i.e. the noise distribution). It follows a highly symmetrical distribution around zero, based on which we define a threshold to remove noise from our final maps (threshold set by the red vertical line, the 95$^{th}$ percentile of the negative noise distribution).}
\label{fig:histonoise}
\end{center}
\end{figure}

We define the threshold relative to the map of propagated uncertainties, with the additional simple assumption that 
the noise randomly introduced by
ICA is symmetric about zero. In principle, there is no mechanism
within ICA that should asymmetrically bias the noise toward positive values, and this is confirmed by the histograms of the flux distribution
of $s2$ over areas where no dust emission is expected (e.g. in the outermost part of the galaxy). For a randomly selected sample of galaxies, these histograms are indeed very symmetric distributions centered around zero (some are shown in Fig.~\ref{fig:histonoise} for illustrative purposes).

In light of the symmetric distribution of noise around zero, we adopt a variable threshold, which we set to $M$ times the propagated map of uncertainties, where $M$ is defined by the $95^{th}$ percentile in the noise distribution measured from the negative pixels in $s2$ divided by the average propagated noise:

\begin{multline}
 \mathrm{flux~threshold} = \left(-\frac{\mathrm{95^{th}~percentile~of~negatives}}{\mathrm{average~propagated~noise}}\right)
 \\
  \times \mathrm{map~of~propagated~noise}.
\end{multline}
As we assume that the distribution of negative pixels is representative of the global noise distribution, such that a symmetrical cut in the 95$^{th}$ percentile 
toward positive values removes all negative noise and also 95\% of the artificially introduced positive noise. What is left is primarily the significant emission in $s2$.  

The fluxes at these threshold locations are returned back to the map $s1$, which, as a result, once again becomes identical to the original image for those positions. We note that some genuinely
identified low-level dust emission may be returned to the stellar map. (At low flux values, where the distributions of genuine dust emission and noise overlap, there is no way to uniquely determine their true contributions.)
In effect, we modify 
the picture of the old stellar light map supplied by the original image by removing flux only if we are confident that it corresponds to dust emission.  This prevents the introduction of additional structure by artificially removing flux consistent with noise.

\subsection{Choosing the best solution}

Because ICA optimally separates two sources in two images, the second iteration, for which the number of sources has been reduced, supplies in principle our best solution. In practice, this translates into a dust color which becomes less red after the second iteration, as a combination of hot dust, evolved star regions and possibly red nuclei have been masked, to converge on a color representative of the spatially dominant diffuse dust emission.

However, when the second iteration returns solutions that are unphysical (i.e. the ICA colors of one or both sources fall outside the range of colors expected for old stars or dust emission, or the identified dust color turns redder after the second iteration), then we take this as an indication that the solution has been biased by the masking of noisy pixels rather than by a true additional secondary source. The solution from the first iteration is chosen in this case (34\% of the cases).

\subsection{Solution postprocessing}
\label{subsec:solpostpro}

Independently of the choice of solution, \textit{a posteriori} processing is necessary, as the light from old stars will be over- or under-estimated in pixels containing dust properties that deviate from the dominant ones.  
There are two cases when we choose to adjust \textit{a posteriori} the information in map $s1$. In the first case, when pixels contain dust with redder colors than the ICA $[3.6]-[4.5]\vert_{s2}$ color and these are unmasked, the solution overestimates the true secondary flux in these pixels. This manifests itself in oversubtractions on the stellar flux map, which we find  to be an acceptable price to pay for the high quality with which a larger fraction of the dust emission can be described with ICA. In these cases we opt to linearly interpolate over the regions of oversubtraction, which are typically small, effectively filling them in with information about the old stellar light from neighboring pixels.   

A second case arises when those red regions have been masked in the second iteration (i.e. regions with the reddest dust emission). Since we do not want to leave that extra flux in the stellar map, we also linearly interpolate over all those masked regions (since they are typically of the order of a few pixels). The map of dust emission is then adjusted in this case, by taking the difference between the original image and the adjusted old stellar map. 
Naturally, the colors of the sources are also recalculated after these slight flux modifications.

\subsection{Quantifying uncertainties}
\label{subsect:uncertainties}
Several uncertainties combine to set the level of accuracy that we can obtain in our maps of the distribution of old stellar light, due to both systematic and measurement errors. For clarity, Table~\ref{table:errors} summarizes the typical values of the different uncertainties involved.

The 3.6 and 4.5\,$\mu m$ images on which we
base our separation have photometric uncertainties, which propagate through the ICA method into the final maps we produce. The original photometric uncertainties are quantified via
the sigma maps ($\sigma_1$, $\sigma_2$) that we obtain from the S$^{4}$G  weight maps and, according to eq. (1), they propagate in to the following uncertainties:

\begin{multline}
 \sigma_{11}^2 = 
 \frac{
 \mathrm{ZP}^2
 }{
 10^{0.8(cs1)}-10^{0.8(cs2)}
 } \times
 \\ 
  \left[(\sigma_1^2 + \Delta \mathrm{sky_1}^2) 
   + \frac{10^{0.8(cs2)} }{\mathrm{ZP}^2} (\sigma_2^2 + \Delta \mathrm{sky_2}^2) \right]
 ,
 \end{multline}

and

 \begin{multline}
  \sigma_{12}^2 = 
 \frac{
 \mathrm{ZP}^2
 }{
 10^{0.8(cs2)}-10^{0.8(cs1)}
 } \times
 \\ 
  \left[(\sigma_1^2 + \Delta \mathrm{sky_1}^2)
   + \frac{10^{0.8(cs1)} }{\mathrm{ZP}^2} (\sigma_2^2 + \Delta \mathrm{sky_2}^2) \right]
 ,
 \end{multline}
 
 \noindent
where $\mathrm{ZP}=280.9/179.7$, $cs1 = [3.6]-[4.5]\vert_{s1}$, $cs2 = [3.6]-[4.5]\vert_{s2}$, and $\Delta \mathrm{sky_1}$ and $\Delta \mathrm{sky_2}$ refer to the uncertainties in the determination of the sky for a given galaxy in each of the bands. In particular, there are two sky uncertainties computed within S$^{4}$G Pipeline~3: one reflects Poisson noise, while the other refers to the large-scale background errors 
\citep[rms calculated within and among different sky boxes;][see also \citet{2009ApJ...703.1569M} for a detailed discussion]{2013ApJ...771...59M}.
Our global estimate of the sky error is calculated as $\Delta \mathrm{sky}=\sqrt{\Delta \mathrm{sky_{Poisson}^2} + \Delta \mathrm{sky_{large~scale}^2}}$, which has a typical (median) value of 0.013\,MJy/sr for both bands. Here, we neglect any uncertainties associated to the photometric zero points.

There is also uncertainty intrinsic to the ICA method, namely the reliability or uniqueness of a solution identified in any given measurement set.  
Our algorithm includes a quantification of the uniqueness of the solution by  performing ICA on each measurement set $N$ times, each time with a different initial seed matrix of mixing coefficients (bootstrapping).  Perturbations to the initial seed are fixed for all galaxies in the sample and represent $N$ possible realistic mixtures of old stars and dust emission. Optimally, ICA quickly converges to its final solution independent of the initial seed \citep{2012ApJ...744...17M}. The range of final mixing coefficients sets the uncertainty on the final average solution, and this ICA color uncertainty corresponds typically to values of order $\sim 0.07$ mag. The ICA tests presented in the Appendices confirm the meaningfulness of these uncertainties.

The $s1$ color change associated with the second iteration is typically of the order of $\sim 0.04$ mag. Interestingly, in 88\% of the cases, this color change is well constrained by the initial ICA bootstrap uncertainty (i.e. the change occurs within the $s1$ color error bar). Taking this into account, and considering the typical values listed in Table~\ref{table:errors}, we can state that the ICA boostrap error is the dominant source of uncertainty, and provides a good estimate of how accurate our solutions are (see Appendix~B).

\begin{table*}[t]
\begin{center}
\caption{Uncertainties involved in the pipeline.\tablenotemark{\it a} }
\begin{tabular}{cc}
\tableline\tableline
 Uncertainty & Typical value (mag)\\   
\tableline
 Error of original global color & $ 0.001 $ \\
 ICA bootstrapping error & $0.074$     \\
 Propagated photometric error $[3.6] - [4.5]\vert_{s1}$ & $0.028$     \\
 $2^{nd}$ iteration change on $[3.6] - [4.5]\vert_{s1}$ & $0.036$ \\
\tableline
\end{tabular}
\label{table:errors}
\end{center}
\tablenotetext{a}{Typical values understood as the median values of the distributions.}
\end{table*}

\subsection{Quality flagging}
\label{subsec:qflags}

To account for the varying quality of solutions, flags are provided along with the final products.
Flagging has been independently performed by two of the authors (Sharon Meidt, Miguel Querejeta) and an external classifier (Emer Brady), according to well-defined criteria: 
\begin{enumerate}
 \item Is the physical distribution 
of the identified dust in accordance with the signatures that appear in the color map?
\item Are there significant oversubtractions in the map of old stellar flux?
\item Are there any artifacts that prevent ICA from obtaining a correct solution?
\end{enumerate}

Three quality flags have been established: 1--\textit{excellent}, 2--\textit{acceptable}, 3--\textit{bad}, 
and the statistical mode is chosen as the final classification (cases in which all three classifiers disagree have been revised
and individually discussed).
Depending on the specific application of the mass maps, either only those galaxies classified as 1 or both 1 and 2 will be suitable; also, some of the solutions
classified as 3--\textit{bad} correspond to galaxies in which the input data was contaminated by artifacts (e.g. mux bleeds or saturated PSFs), and
a personalized case-by-case treatment of the masks can potentially improve the quality of those solutions (but we have not done it to avoid introducing a subjective component, and for homogeneity of the pipeline results).

It should be noted that, within the group of more than 1200 galaxies that constitute our final {\it science sample} (Sect.~\ref{sec:icasources}), very few have the {\it bad} quality flag, `3' (only 3.5\%). This is, to a great extent, because we have conservatively excluded from the analysis all galaxies with low S/N, which systematically lead to solutions of poorer quality.

\section{Stellar mass maps and products released}
\label{subsec:massmapsproducts}

The dust-free flux maps that we produce with our Pipeline~5 can be directly converted into mass maps by choosing the appropriate mass-to-light ratio and assumed distance. In Sect.~\ref{subsec:massmaps} we refer to a possible strategy to choose $M/L$, including the necessary conversions into appropriate units. In Sect.~\ref{subsec:products} the product release of S$^{4}$G Pipeline~5 is described.

\subsection{Converting to stellar mass maps}
\label{subsec:massmaps}

Using the IRAC zero magnitude flux density at 3.6 
and 4.5\,$\mu m$, $\mathrm{ZP}_{3.6\mu m}=280.9$\,Jy and $\mathrm{ZP}_{4.5\mu m}=179.7$\,Jy
 \citep[][]{2005PASP..117..978R}, and the corresponding absolute magnitude of the Sun, $M_{\odot}^{3.6}=3.24$\,mag \citep{2008AJ....136.2761O}, we obtain the following relationship for the 3.6\,$\mu m$ IRAC band:

\begin{equation}
1 \,\mathrm{MJy/sr} = 704.04 \,L_{\odot}/\mathrm{pc}^2.
\end{equation}

\noindent

Starting from an ICA-corrected flux density measurement $S_{3.6 \mu m}$ in the \textit{Spitzer} units of MJy/sr, the \textit{stellar} mass contained by a pixel (0.75'') can be obtained as:

\begin{equation}
\frac{M}{M_{\odot}} = 9308.23 \times \left(\frac{S_{3.6 \mu m}}{\mathrm{MJy\,/\,sr}}\right) \times \left(\frac{D}{\mathrm{Mpc}}\right)^2 \times \left(\frac{M/L_{3.6 \mu m}}{M_{\odot}/L_{\odot}}\right).
\end{equation}

For a detailed discussion on how to choose the optimal 3.6\,$\mu m$ mass-to-light ratio, $M/L_{3.6 \mu m}$, the reader is referred to \citet{2014ApJ...788..144M} and \citet{2014ApJ...797...55N}.
Here we adopt a single M/L=0.6 (assuming a Chabrier IMF), which according to both sets of authors can convert the 3.6\,$\mu m$ old stellar flux (with dust removed) into stellar mass with an accuracy of $\sim 0.1$~dex. Given that the dependence on age and metallicity of the M/L at 3.6\,$\mu m$ is relatively small for old stellar populations, \citet{2014ApJ...788..144M} advocate for this constant M/L and its uncertainty assuming a universal age-metallicity relation, together with the constraint on metallicity (and thus age) provided by the [3.6]-[4.5] color. Following an independent, empirical strategy, \citet{2014ApJ...797...55N} argue for a comparable value, without invoking such an argument. A $M/L$ of 0.6 is also found to be representative in stellar population synthesis models, extended to the wavelength range of 2.5--5\,$\mu m$ using empirical stellar spectra by \citet{2015MNRAS.449.2853R}.

\subsection{Public data products}
\label{subsec:products}

To give users the opportunity to choose their preferred distance and $M/L$, we release the map of the old stellar flux (map $s1$), along with the dust map ($s2$). To allow for the choice of a spatially varying $M/L$, we also provide a color map for the old stars. This color map conserves the original [3.6]-[4.5] color in all pixels without non-stellar emission (i.e., where $s2$=0 after thresholding), and  $[3.6]-[4.5]\vert_{s1}$ elsewhere. Additionally, the quality flags, colors and integrated fluxes of each component are made public in a table format.

Our recommended post-processing strategy is explained in Sect.~\ref{subsec:solpostpro}, which includes interpolation over masked areas. However, a myriad of interpolation techniques exist, and some users may even prefer to leave all the original flux unchanged for those regions. Therefore, we also provide the masks used,
making it possible for different strategies to be applied. In a final step, for aesthetical purposes, dust maps have been smoothed in the areas of significant dust emission using a Gaussian kernel
of $\sigma = 3 \mathrm{px}$ and conserving total flux between $s1$ and $s2$, but unsmoothed maps are available upon request. The release of Pipeline~5 products takes place on the NASA/IPAC Infrared Science Archive (IRSA).

\section{Final sample with ICA solutions}
\label{sec:icasources}

In Sect.~\ref{applyica} we introduced an initial cut in signal-to-noise to make sure that we apply ICA in a regime in which it can perform correctly. Excluding all galaxies with $S/N<10$ made 667 objects be initially discarded. Additionally, it was explained in Sect.~\ref{subsec:icaneed} that 376 galaxies from S$^4$G (16.3\% of total) have original colors which are already compatible with an old stellar population. On those objects, and using the weighted mean of the original color to discriminate, we do not apply ICA, given that the fractional contamination from dust is low (typically below $\sim 15\%$). We have shown (see Appendix~A) that uncertainties become large under those conditions, and we risk incurring a larger error by inaccurately removing the dust emission present, if any, than not correcting for it.

In fact, the group of 376 galaxies with original blue colors to which we do not apply ICA is clearly dominated by early-type galaxies. The majority (201 galaxies) are ellipticals or S0s, and the rest are predominantly early-type spirals. For those early-type galaxies, based on scaling relations from longer wavelengths, we can assume a maximum fraction of flux due to dust of $\sim 15 \%$, which constrains the maximum global error due to not applying a dust correction (see Appendix~A).

Once galaxies with low signal-to-noise and original blue colors have been excluded, the optimized algorithm explained in Sect.~4 has been applied to the rest of the S$^4$G sample, which includes galaxies across the whole
Hubble sequence, covering a wide range of masses and star-formation rates. This extends the work by \citet{2012ApJ...744...17M} to a much broader range of galaxy types and observational characteristics. Table~\ref{tab-colors} summarizes the different groups of galaxies that we have just mentioned.

\begin{table*}[t]
\begin{center}
\caption{Fraction of S$^4$G Galaxies to Which ICA has been Applied. \label{tab-colors}}
\begin{tabular}{lccc}
\tableline\tableline
 Group/Class & Number of Galaxies & Fraction\tablenotemark{a} & Criterion\\
\tableline
 All S$^4$G & 2352	& ---	& --- \\
 Good P3 data\tablenotemark{a} & 2308 & 100\%	& --- \\
 Low S/N: discarded & 644 & 28.0\% & $S/N < 10$ \\
 Little dust: ICA not applied & 376 & 16.3\% & $[3.6]-[4.5]\vert_{\mathrm{orig,weighted}} < 0$ \\
 ICA applied & 1288 & 55.8\% & $[3.6]-[4.5]\vert_{\mathrm{orig,weighted}} > 0$ \\
 $\bullet$ $s1$ compatible with old stars & 1251 & 54.2\% & $-0.2 < [3.6]-[4.5]\vert_{s1} < 0$ \\
 $\bullet$ $s1$ incompatible with old stars & 37 & 1.6\% & rest \\
\tableline
 Total good mass maps & 1627 & 70.5\% &  --- \\
\tableline
\end{tabular}
\end{center}
\tablenotetext{a}{At the time of running our pipeline, not all of S$^4$G galaxies had Pipeline 3 (P3) data available; the ones that failed at P3 are typically those with the poorest data quality, and even if some have been added later on, we prefer not to include them in the analysis, as they are few and likely problematic. The {\it fraction of total} is therefore referred to the 2308 galaxies with good P3 data at the moment we ran this Pipeline 5.}
\end{table*}

We find that, after applying ICA, 1251 galaxies have $s1$ colors compatible with those expected for old stars, $-0.2<[3.6]-[4.5]\vert_{s1}<0$. The former range matches the colors expected for K and M giants \citep[see \citet{2014ApJ...788..144M}, and as observed in old, dust-free E/S0 galaxies,][and in globular clusters, \citet{2014ApJ...797...55N}]{2012MNRAS.419.2031P}. Only 37 galaxies have final $s1$ colors that do not match the range expected for old stars. Adding the (mainly early-type) galaxies to which we did not apply ICA because they already had original colors compatible with old  stars, this means that we have a total of more than 1600 galaxies with good mass maps.

\begin{figure*}[t!]
\begin{center}$
 \begin{array}{ccc}
\includegraphics[width=0.28\textwidth]{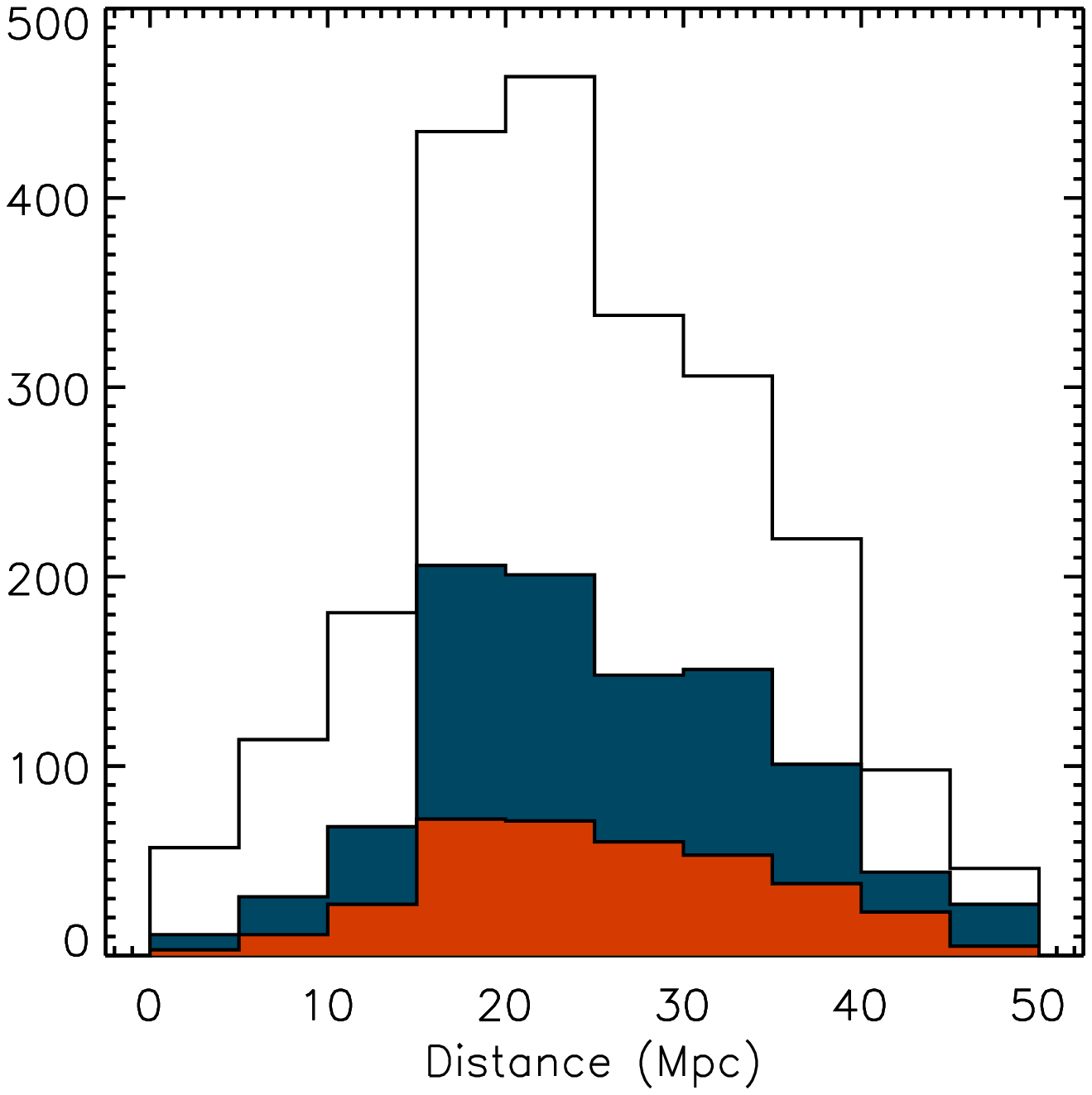} &
\includegraphics[width=0.28\textwidth]{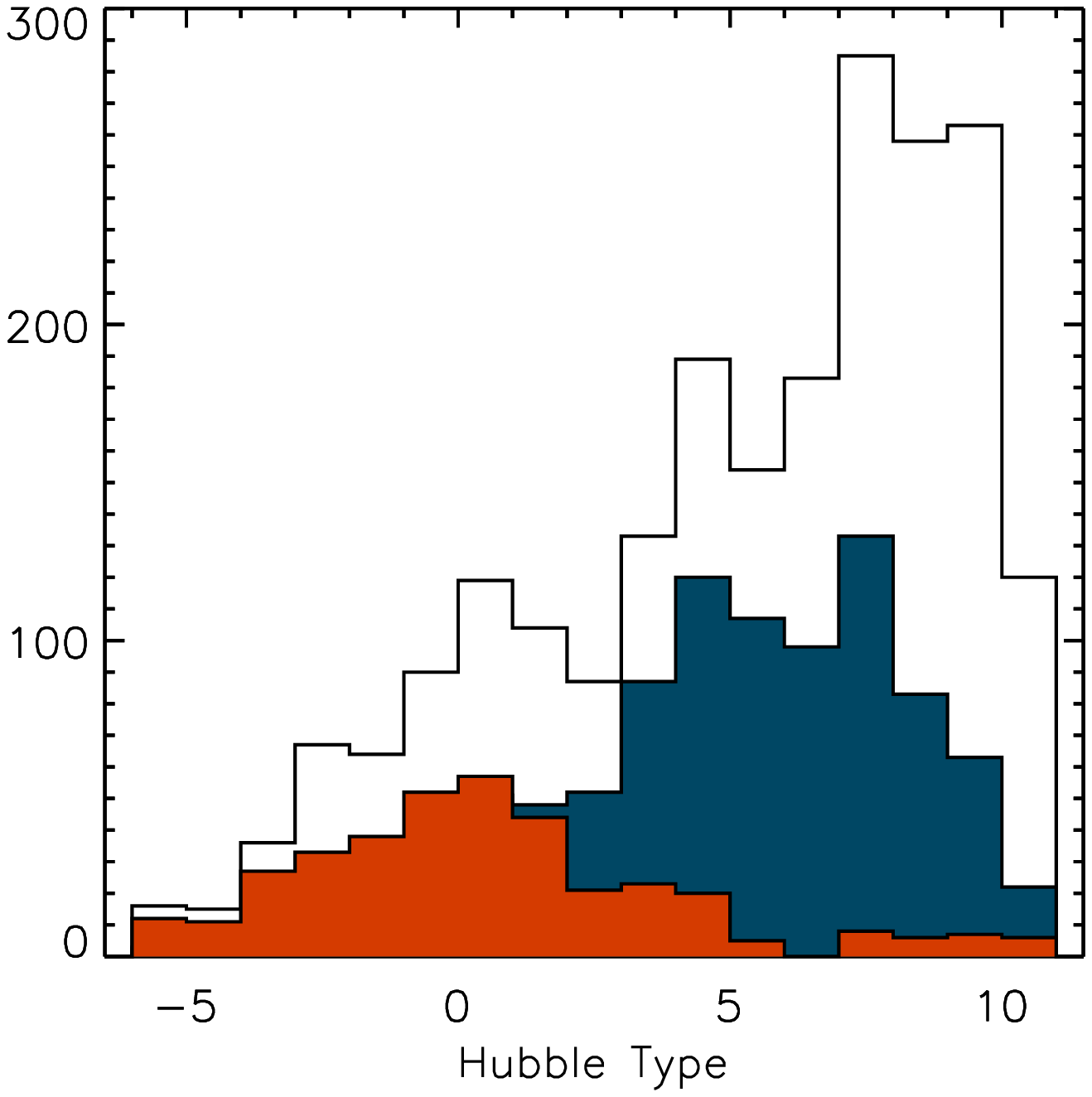} &
\includegraphics[width=0.28\textwidth]{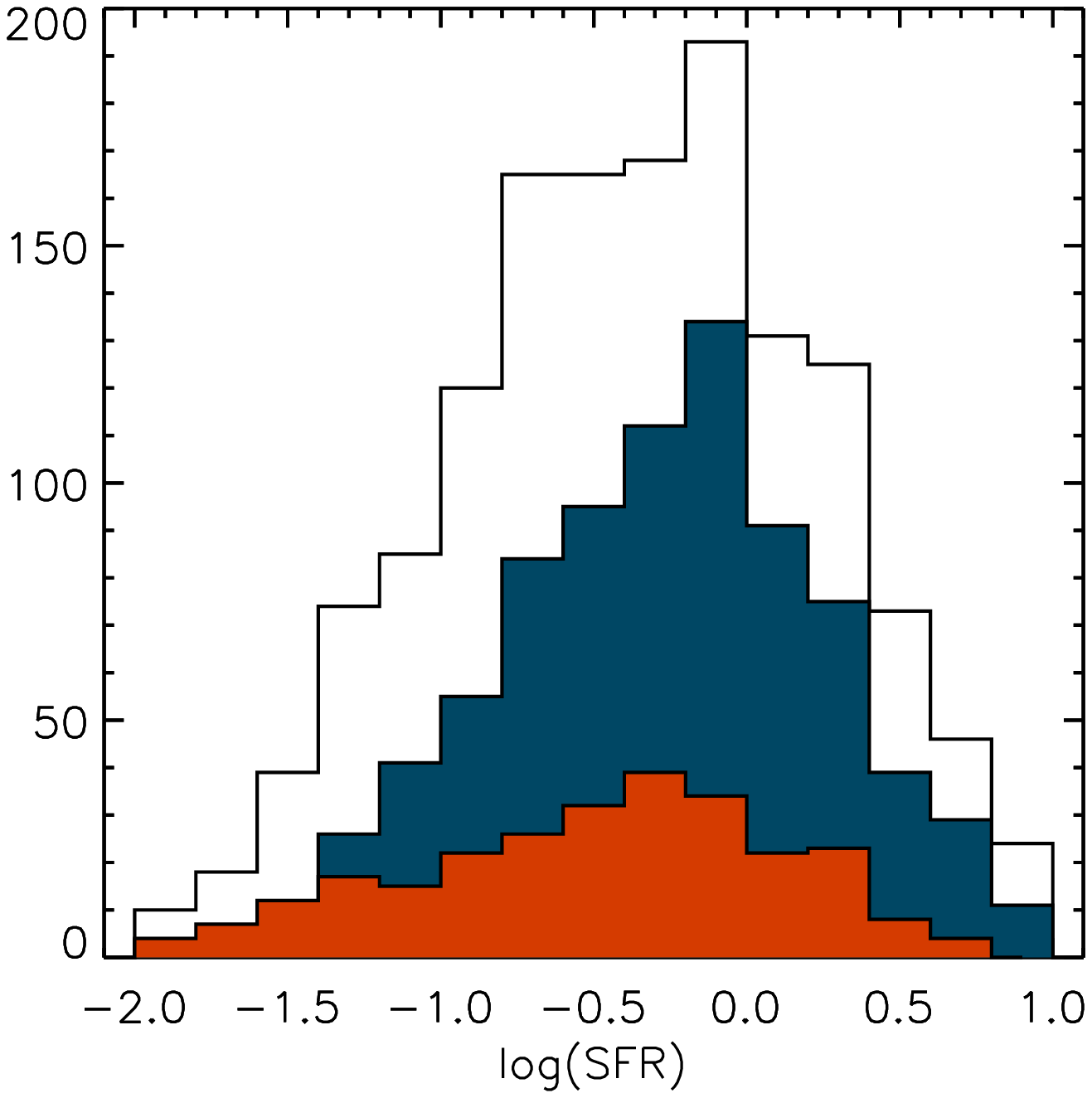}
\end{array}$
 \caption{Histograms showing the distances, morphological types and star formation rates spanned by the galaxies in our science sample (blue). Overplotted in red is the distribution of the galaxies to which ICA was not applied (compatible with old stars). The global S$^4$G distributions are shown by the empty black histograms.}
\label{fig:histograms_sample1}
\end{center}
\end{figure*}

Fig.~\ref{fig:histograms_sample1} shows the distribution of the galaxies in our final ICA-corrected catalog according to distance, morphological type and SFRs, proving that it covers a representative range in all three properties. For reference, the distribution of galaxies to which we did not apply ICA due to original colors compatible with old stars is overlaid in red, and an empty black line shows the global distributions for the entire S$^4$G.

\section{Results}
\label{sec:results}

\subsection{Trends with Hubble type} 

\begin{figure}[t!]
\begin{center}
 \includegraphics[width=0.48\textwidth]{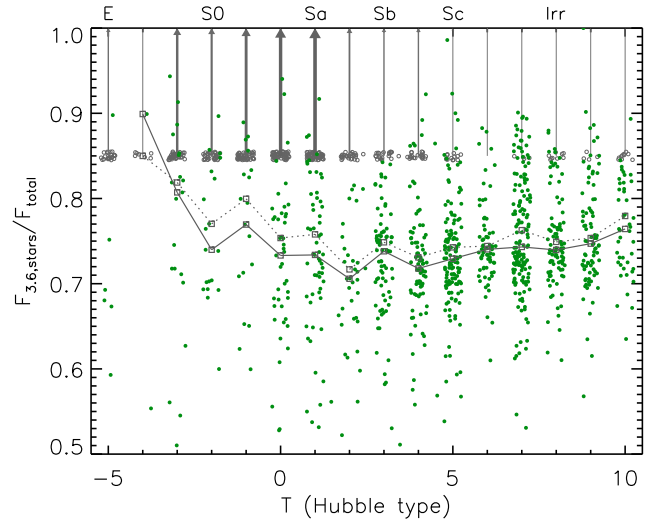}
 \caption{Fraction of total observed 3.6\,$\mu m$ flux arising from old stars as a function of Hubble type. The open circles located at $F_{stars}/F_{total} \approx 0.85$ correspond to the galaxies with original blue colors to which ICA was not applied; they are included here as lower limits for reference (maximum expected contamination of 15\%, see Appendix~A; thickness and darkness of arrows are proportional to the number of galaxies in each bin). We see that contamination becomes more severe towards later $T$, tracking an increase in dust content until the latest types, where dust and PAHs are fewer. The gray squares joined by a continuous line represent the running medians in each bin, excluding lower limits, while the squares joined by a dotted line are the equivalent considering the lower limits.}
\label{fig:PPR9_fracs2_T}
\end{center}
\end{figure}

Here we explore how the stellar flux fraction ($s1$/total) changes as a function of Hubble type (Fig.~\ref{fig:PPR9_fracs2_T}). We can see that the observed 3.6\,$\mu m$ flux becomes a poorer tracer of stellar mass as the Hubble type increases, because dust emission becomes more significant.
This contamination from dust increases slightly toward late-type galaxies, as shown by the running medians, and seems to decrease again for the latest Hubble types.
The reason for this final decline in contamination is probably due to the fact that very late-type galaxies tend to have less dust as they have lower metallicity. For standard spiral galaxies (Sa-Sc), the stellar component contributes $\sim 70-80 \%$ of the flux (i.e. as much as 20-30\% is coming from dust), with both upper and lower outliers. Moreover, these are only global estimates, and these values can be significantly higher in individual star-forming regions, as shown by \citet{2012ApJ...744...17M}. To account for completeness, we include as lower limits the galaxies to which ICA was not applied due to original blue colors (compatible with old stars, which implies little dust emission): they are shown as grey arrows marking the upper limit of 15\% contamination that is estimated in the Appendix~A (and the arrow thickness is proportional to the number of galaxies per bin).

\subsection{Trends with SFR} 

\begin{figure}[h!]
\begin{center}
 \includegraphics[width=0.48\textwidth]{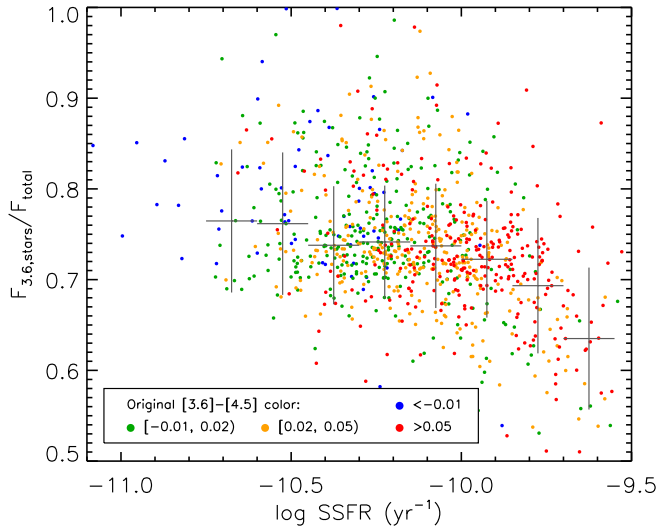}
 \caption{Fractional contribution from old stars to total $3.6 \mu m$ flux as a function of specific star formation rate: there is a clear trend between fractional $s1$ and SSFR, as it can be seen by the almost monotonic drop of the running mean (the width of the running median is the bin size, and height measures the standard deviation of points in each bin). The points represent all galaxies with ICA solutions compatible with old stars and SFR measurements available (a total of 819 galaxies).
 }
\label{fig:P3_fracs1_SFR}
\end{center}
\end{figure}

In an attempt to investigate the main driver behind the varying flux fraction due to old stars at 3.6\,$\mu m$, in Fig.~\ref{fig:P3_fracs1_SFR} we plot that same quantity ($F_{3.6, \mathrm{stars}}/F_\mathrm{total}$) as a function of the specific star formation rates (SSFRs), and find a declining correlation. We calculate SSFRs as the star formation rates divided by our stellar mass, that is derived from the 3.6\,$\mu m$ flux corrected for dust with a M/L=0.6, and integrated within the 25.5 mag/arcsec$^2$ isophote. The SFRs come from IRAS 60 and 100\,$\mu m$ fluxes (the weighted average of the values reported in NED), following the recipe from \citet{2000A&A...354..836L}. Therefore, the subsample shown in Fig.~\ref{fig:P3_fracs1_SFR} is all S$^4$G galaxies with ICA solutions compatible with old stars (within their uncertainties) and with availability of SFRs, which makes a total of 819 galaxies. Galaxies with higher SSFRs have on average higher fractional contamination from dust, which makes the fractional flux due to stars drop (see running median in Fig.~\ref{fig:P3_fracs1_SFR}).

Finally, there is no obvious trend of dust emission fraction with SFR alone: the star formation rate relative to the total mass of the galaxy (SSFR) is a better first-order indicator of how much flux is arising from dust and not from old stars at 3.6\,$\mu m$.

\subsection{Global stellar mass estimates at 3.6\,$\mu m$}
\label{massestimates}

We have successfully isolated for old stellar emission the 3.6\,$\mu m$ images of more than 1600 galaxies in S$^4$G, which provides us with an unprecedentedly large sample of stellar mass estimates for nearby galaxies. This statistically powerful tool allows us to produce an empirical calibration of the mass-to-light ratio (M/L) and its scatter as a function of observed original color of the galaxy.

\begin{figure}[h!]
\begin{center}
 \includegraphics[width=0.48\textwidth]{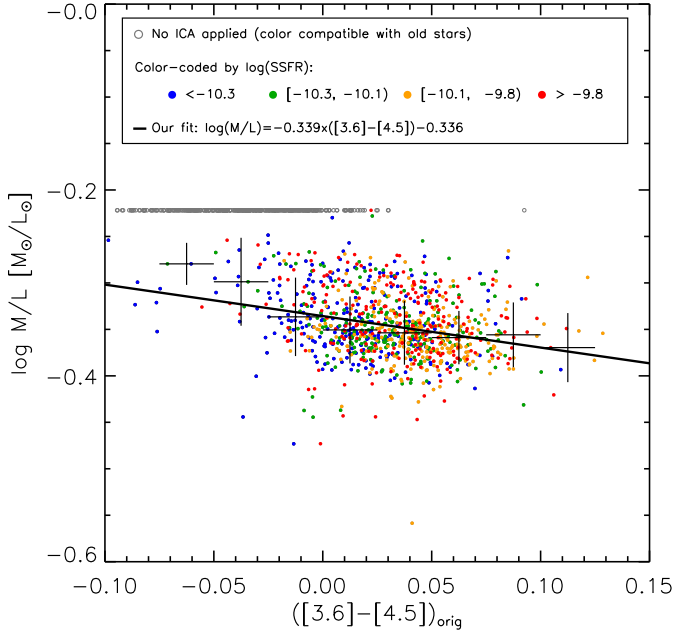}
 \caption{Our empirical mass-to-light ratio is obtained by multiplying the total flux in our dust-free maps by the M/L$_{3.6 \mu m}$=0.6 ($M_{\odot}/L_{\odot}$) suggested by \citet{2014ApJ...788..144M}, and dividing it by the total original $3.6 \mu m$ flux inside the ICA solution area. 
 This is plotted against the original [3.6]-[4.5] color (obtained from the integrated 3.6 and 4.5\,$\mu m$ fluxes within our area of analysis, which makes this differ from the weighted mean color that we used to discriminate to which galaxies we apply ICA).  The galaxies with original negative weighted mean colors to which we did not apply ICA are shown with gray open circles for reference, with a constant M/L$_{3.6 \mu m}$=0.6 ($M_{\odot}/L_{\odot}$).  The color-coding shows that there is some trend with SSFR, as it could be expected from Fig.~\ref{fig:P3_fracs1_SFR}: it is the dustiest galaxies with highest SSFRs that most significantly diverge from the constant value of M/L=0.6. Running medians are shown on top, in color intervals of 0.025 (the vertical length is the standard deviation of points in each bin). The linear fit is obtained using \texttt{FITEXY} in the range $-0.1<[3.6]-[4.5]<0.15$.
 }
\label{fig:PPR1_MoverLa}
\end{center}
\end{figure}

Our empirical {\it effective} mass-to-light ratio as a function of [3.6]-[4.5] global color is shown in Fig.~\ref{fig:PPR1_MoverLa}. This is the mass-to-light ratio required to convert the uncorrected 3.6 micron flux --including both stellar and non-stellar emission-- into stellar mass, calculated by dividing the stellar mass implied by the $s1$ map for each galaxy by the total original luminosity in the analysis area.  The stellar mass is calculated by multiplying the total luminosity in the $s1$ map by a constant mass to light ratio 0.6\,$M_\odot / L_\odot$ \citep{2014ApJ...788..144M}, and divided by the total original luminosity within our area of analysis. It is important to note here that the abscissa is the original integrated color that would be available, for instance, to an observer that cannot spatially resolve a galaxy; this is different from the weighted mean color (based on the pixel-to-pixel photometric uncertainty) that we used to discriminate which galaxies to apply ICA to in the first step of our pipeline.

Overplotted in Fig.~\ref{fig:PPR1_MoverLa} as a thick, continuous  line is the regression to the data points in the range $-0.1<[3.6]-[4.5]<0.15$, which provides the calibration:

\begin{equation}
 \log(M/L)=-0.339 (\pm 0.057) \times ([3.6]-[4.5]) -0.336 (\pm 0.002).
\end{equation}

We can also express this in the following equivalent form, which provides stellar mass in terms of the measured 3.6 and 4.5\,\,$\mu m$ flux densities (expressed in Jy) and the distance (in Mpc):

\begin{equation}
 \frac{M_*}{M_{\odot}}=  10^{8.35} \times \left(\frac{F_{3.6}}{\mathrm{Jy}}\right)^{1.85} \times \left(\frac{F_{4.5}}{\mathrm{Jy}}\right)^{-0.85} \times \left(\frac{D}{\mathrm{Mpc}}\right)^{2}.
\end{equation}

The decreasing trend in Fig.~\ref{fig:PPR1_MoverLa} suggests that color is indeed a good first-order indicator of the level of contamination and, therefore, of the M/L for a given galaxy. The impact of contamination on the M/L is demonstrated by the color-coding based on SSFRs: since high specific star formation rates are typically associated with redder galaxies (see color coding in Fig.~\ref{fig:P3_fracs1_SFR}), this provides a physical connection to the trend found in Fig.~\ref{fig:PPR1_MoverLa}. Galaxies in which star formation is relatively more prominent (higher SSFRs) have more dust emission in these bands, leading to redder colors and thus also requiring a lower M/L to convert the light into mass.

While Fig.~\ref{fig:PPR1_MoverLa} suggests that the color can be used to constrain the \textit{effective} M/L, there is still considerable scatter at fixed [3.6]-[4.5] color.  We therefore recommend adopting an uncertainty of $\sim$0.2\,dex on our empirical relation, which also accounts for the 0.1\,dex uncertainty associated with our adopted stellar M/L (see Sect.~\ref{subsec:massmaps}). (We note, however, that a method that can correctly remove dust emission, such as the one that we have presented here, should be the preferred approach for esimating stellar masses.)

\begin{figure}[h!]
\begin{center}
 \includegraphics[width=0.48\textwidth]{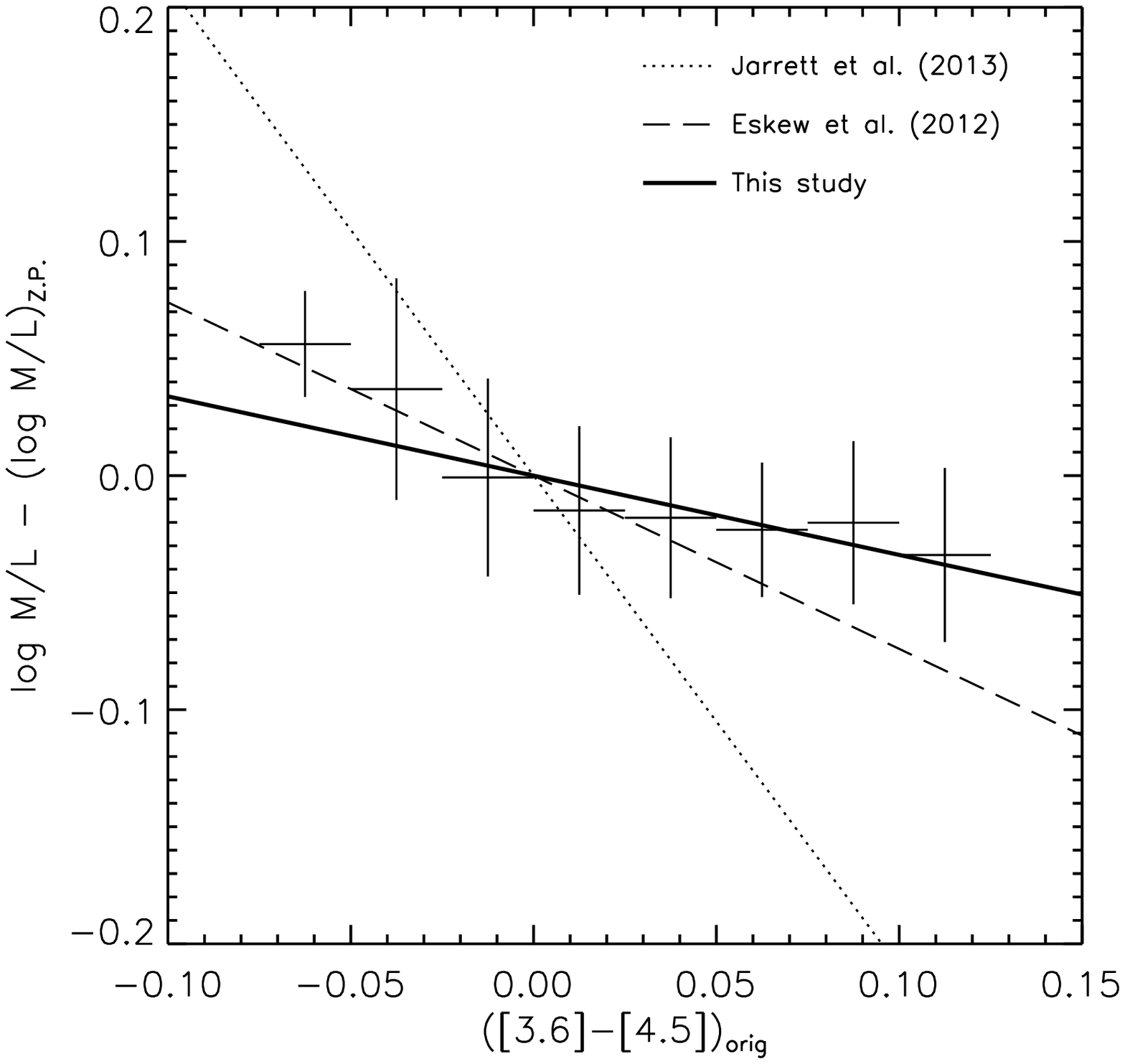}
 \caption{The linear fit to the effective M/L measured after removing dust emission from the S$^4$G galaxies is plotted, along with the running medians, relative to the purely stellar M/L$_{3.6 \mu m}$=0.6 ($M_{\odot}/L_{\odot}$) from \citet{2014ApJ...788..144M}. Here, we compare against the trend found by \citet{2013AJ....145....6J}, which is shown with a dotted line,
 and the color-dependent M/L$_{3.6 \mu m}$ from \citet{2012AJ....143..139E}, with a dashed line. For this comparison, all relations have been normalized so that log(M/L)=0 at [3.6]-[4.5]=0. }
\label{fig:PPR1_MoverLb}
\end{center}
\end{figure}

Our relation between the effective M/L and [3.6]-[4.5] color, calibrated using our optimal stellar mass estimates, compares well with other relations in the literature. As part of the WISE Enhanced Resolution Galaxy Atlas, \citet{2013AJ....145....6J} also found a declining correlation between W1-W2 color (the 3.4 and 4.6\,$\mu m$ WISE bands) and M/L at 3.6\,$\mu m$, based on their observational stellar mass estimates of a few ($\sim 15$) galaxies. Fig.~\ref{fig:PPR1_MoverLb} shows that the slope that we find is considerably shallower, based on a sample nearly 100 times larger.
In this plot we show $\Delta M/L$ in order to highlight the IMF-independent slopes of each relation and avoid differences in vertical offset due to choice of the IMF.

We also find good agreement with the stellar M/L estimated by \citet{2012AJ....143..139E}. Using a different method, based on spatially resolved star formation histories (SFHs) in the Large Magellanic Cloud (LMC), they calibrated a conversion between 3.6 and 4.5\,$\mu m$ fluxes and stellar mass. This can be translated into an equivalent color-dependent M/L ($\log M/L = -0.74 ([3.6]-[4.5]) - 0.236$), which is overplotted on Fig.~\ref{fig:PPR1_MoverLb}, also normalized so that log(M/L)=0 at [3.6]-[4.5]=0.
It is remarkable that, in spite of the very different method used, their declining slope is very similar to the fit to our data. Finally, it is also worth mentioning that \citet{2014AJ....148...77M} recently point to a constant $M/L_{3.6 \mu m} = 0.47$, with a scatter of 0.1~dex, which is also compatible with our result.

\section{Summary and Conclusions}

We have developed a pipeline to obtain maps of the flux from old stars for the 3.6\,$\mu m$ images in the {\it Spitzer} Survey of Stellar Structure in Galaxies (S$^4$G). Following two different approaches, \citet{2014ApJ...788..144M} and \citet{2014ApJ...797...55N} have shown that the mass-to-light ratio for old stars at 3.6\,$\mu m$ varies modestly with the age and metallicity of the population, so that a constant mass-to-light ratio is applicable with an uncertainty of $\sim 0.1$~dex. Therefore, when only old stars contribute to the flux at 3.6\,$\mu m$, a simple re-scaling of the corresponding dust-free S$^4$G image effectively constitutes a stellar mass map. However, this is complicated by the fact that dust emission usually contributes a significant fraction of that flux ($\sim 10-30\%$); therefore, a way to automatically correct for this dust emission is highly desirable.

At the core of our method is the application of an Independent Component Analysis technique (ICA), first presented in \citet{2012ApJ...744...17M}, to remove dust emission from these bands. We simultaneously use the information from the 3.6\,$\mu m$ and 4.5\,$\mu m$ images, which are uniformly available for the whole S$^4$G sample. Old stars (age~$\tau \sim$~2-12~Gyr) have colors in the range $-0.2 < [3.6] - [4.5]\vert_{\mathrm{stars}} < 0$, whereas dust corresponds to $[3.6] - [4.5]\vert_{\mathrm{dust}} > 0$, and this difference in SED allows ICA to separate both components. Our method results in an unbiased view of the flux from old stars, which can then be readily used to chart the stellar mass distribution.

Some galaxies in S$^4$G have original colors which are already compatible with an old stellar population, but they are the exception rather than the rule (376, 16\% of the sample). These are preferentially early-type galaxies, and they have little dust emission: we therefore do not run ICA on them. We also exclude from the ICA analysis all galaxies with average signal-to-noise below $S/N<10$, to make sure that we apply our pipeline in a regime in which ICA can perform correctly. 

For 1251 galaxies (54\% of the overall sample), ICA is able to find a trustworthy separation into dust emission and old stars. For these galaxies, stars typically contribute 70-90\% of the flux at 3.6\,$\mu m$, with dust contamination peaking at $T \sim 5$. This dust contamination shows a strong correlation with specific star formation rates, confirming that the dust emission that we are correcting for is mostly related to star formation.

We have also profited from the statistical power of such a large sample of accurate mass estimates to calibrate a relationship between the observed [3.6]-[4.5] color and the empirical M/L$_{3.6 \mu m}$ that should be applied to obtain the same stellar masses as we measure. Our regression line is $\log(M/L)=-0.339 (\pm 0.057) \times ([3.6]-[4.5]) -0.336 (\pm 0.002)$, for an assumed M/L$_{3.6 \mu m}=0.6~M_\odot/L_\odot$ for the old stellar population.
An equivalent expression to convert 3.6 and 4.5\,$\mu m$ fluxes directly into stellar mass is provided in Sect.~\ref{massestimates}.
The correlation shows a large scatter, however, which points out the necessity to apply a method like the one we have described if higher precision is required, or if one is interested in the spatial distribution of such stellar mass.

In conclusion, we have produced maps which reliably trace the old stellar flux for a large fraction of the S$^4$G sample. These maps, which trace the distribution of stellar mass, will be made publicly available through IRSA (along with the S$^4$G archive), providing a powerful tool for the astronomical community. Additionally, we have outlined a strategy to remove dust emission from mid-IR images, and analyzed the requirements for a successful application. This could be relevant for similar studies that aim to exploit {\it Spitzer} archival data, and, eventually, with the advent of the James Webb Space Telescope (JWST), this technique should prove very useful to push spatially resolved galaxy mass estimations further to higher redshifts.

\acknowledgments

We are grateful to the anonymous referee for very insightful comments and suggestions, which have helped us improve the quality of the paper.

We acknowledge financial support to the DAGAL network from the People 
Programme (Marie Curie Actions) of the European Union's Seventh Framework
Programme FP7/2007- 2013/ under REA grant agreement number PITN-GA-2011-289313.

M.Q. acknowledges the International Max Planck Research School for Astronomy
and Cosmic Physics at the University of Heidelberg (IMPRS-HD).
We thank Emer Brady for acting as an external classifier for  assigning quality flags, and Sim\'on D\'iaz-Garc\'ia for helpful comments. K.S., J-C.M-M, and T.K acknowledge support from the National Radio Astronomy
Observatory is a facility of the National Science Foundation operated under
cooperative agreement by Associated Universities, Inc. E.A. and A.B. acknowledge financial support from the CNES (Centre National d'{\'E}tudes Spatiales). LCH acknowledges support from the Kavli Foundation, Peking University, and the Chinese Academy of Science through grant No. XDB09030100 (Emergence of Cosmological Structures) from the Strategic Priority Research Program.

This research has made use of the NASA/IPAC Extragalactic Database (NED) which is operated by the Jet Propulsion Laboratory, California Institute of Technology, under contract with the National Aeronautics and Space Administration.

\appendix
\section{Robustness of ICA solutions for S$^4$G images}
\label{sec:appendix}

In order to quantify the conditions under which ICA retrieves robust
solutions, we have constructed a set of models representing realistic
distributions of stars and dust in nearby galaxies. These models cover a range
of dust colors and fractional dust flux contributions, and also allow for the
inclusion 
of a third component that emulates hot dust in HII regions. We consider the
latter of such models in order to demonstrate the improvement to solutions possible when a second ICA
iteration is applied in the presence of a second non-stellar source (three sources in total; see Sect.~\ref{ICA2}).

\subsection{Input parameters and models}

We conduct our tests by creating 2-dimensional realistic distributions of the emission from old stars and dust.
As input models we utilize the spatial distributions of the components $s1$ (old stellar light) and $s2$ (dust emission)
identified with ICA for four galaxies that cover a range in morphology.
For these specific galaxies, the $s1$ and $s2$ components were shown to be representative of old stars and dust in \citep{2012ApJ...744...17M}. Based on these, our model input maps at 3.6 and 4.5\,$\mu m$ are generated in the following way:
(a) the color of old stars is fixed at [3.6]-[4.5]=-0.12;
(b) the color of dust is varied between [3.6]-[4.5]=0 and [3.6]-[4.5]=0.6;
(c) the fractional contribution of dust emission to the total flux is varied from 5 to 40\%.

In each case, we verified that the resulting colors of each of the combined $s1$+$s2$ models are consistent with the observed colors
in the S$^4$G sample, which cover the range $-0.05<[3.6]-[4.5]<0.20$ (if we exclude the 5\% tail at the upper and lower ends).

\subsection{Controlling parameters}

We run a total of 60 tests per galaxy, covering the total dust color range $0.2 \leq [3.6]-[4.5] \leq 0.7$ in steps of 0.1, and varying the fractional contribution of dust by 10, 20, 30 and 40\%. We also explore the dependence on the
dust spatial distribution by repeating each battery of experiments on four different galaxy models; for reference, models 1 through 4 are based on NGC\,2976, NGC\,3184, NGC\,4321 and NGC\,5194, respectively.

\subsubsection{Relative color difference between s1 and s2}

As explained above, we have fixed the stellar color to $[3.6]-[4.5]\vert_{\mathrm{stars}}=-0.12$.
In
testing we have found that only the {\it relative} difference between the colors
imposed for the dust and stars impacts the colors determined by ICA for a given
model; tests with a stellar color -0.02 return identical results to those
presented below when the 
$[3.6]-[4.5]\vert_{\mathrm{dust}}$ in the model is also shifted by 0.10.

\subsubsection{Fractional contribution of dust emission}
\label{fract_contam}

The maximum fractional contribution of dust to the total flux that we impose, 40\%, is motivated by the fact that we do not expect larger dust contributions in real galaxies. Following \citet{2008MNRAS.389..629B}, we obtain an estimate of the PAH fluxes at 8\,$\mu m$ based on 160\,$\mu m$ measurements from \citet{2012MNRAS.425..763G}. We then calculate the possible range corresponding to those values at 3.6\,$\mu m$ following \citet{2006A&A...453..969F}. That provides the possible dust fluxes at 3.6\,$\mu m$, and for that sample of galaxies, we also obtain the total 3.6 \,$\mu m$ fluxes from NED (as the weighted mean of the available values, using the quoted uncertainties).
This allows us to set an upper limit of 40\% to the fractional contribution of dust flux for star-forming spirals (but that is a rare upper limit; the typical values range 20-25\%). Using the fact that $I_{160 \mu m}$ correlates well with the $f_{dust}/f_{stars}$ parameter, we can make use of the fractions $f_{dust}/f_{stars}$ from \citet{2011ApJ...738...89S} to conclude that the fractional contribution of dust flux for early-type galaxies ($T \le 0$) is a factor of 3 lower than for late-type galaxies, which corresponds to an upper limit of $\approx 15\%$.

In addition, in Sect.~\ref{subsec:icaneed} we introduced our criterion to exclude galaxies with original blue colors from the ICA analysis, as it is an indication that those galaxies are already compatible with an old stellar population. To account for this, in Fig.~\ref{fig:2compB} we identify the models in which that criterion is met with a different color.

\subsection{Findings}

\begin{figure}[h!]
\begin{center}$
 \begin{array}{cc}
\includegraphics[width=0.4\textwidth]{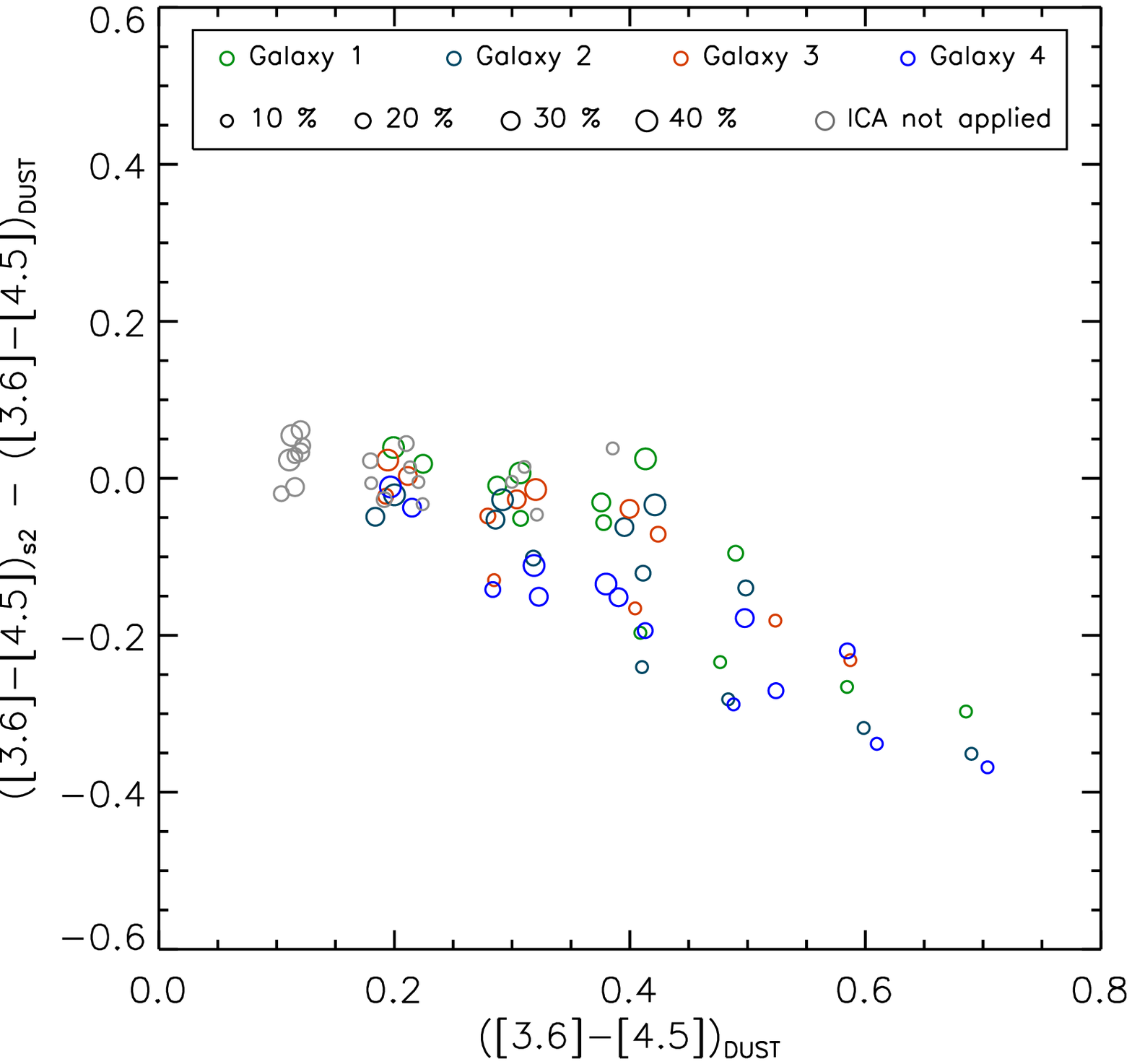} &
\includegraphics[width=0.4\textwidth]{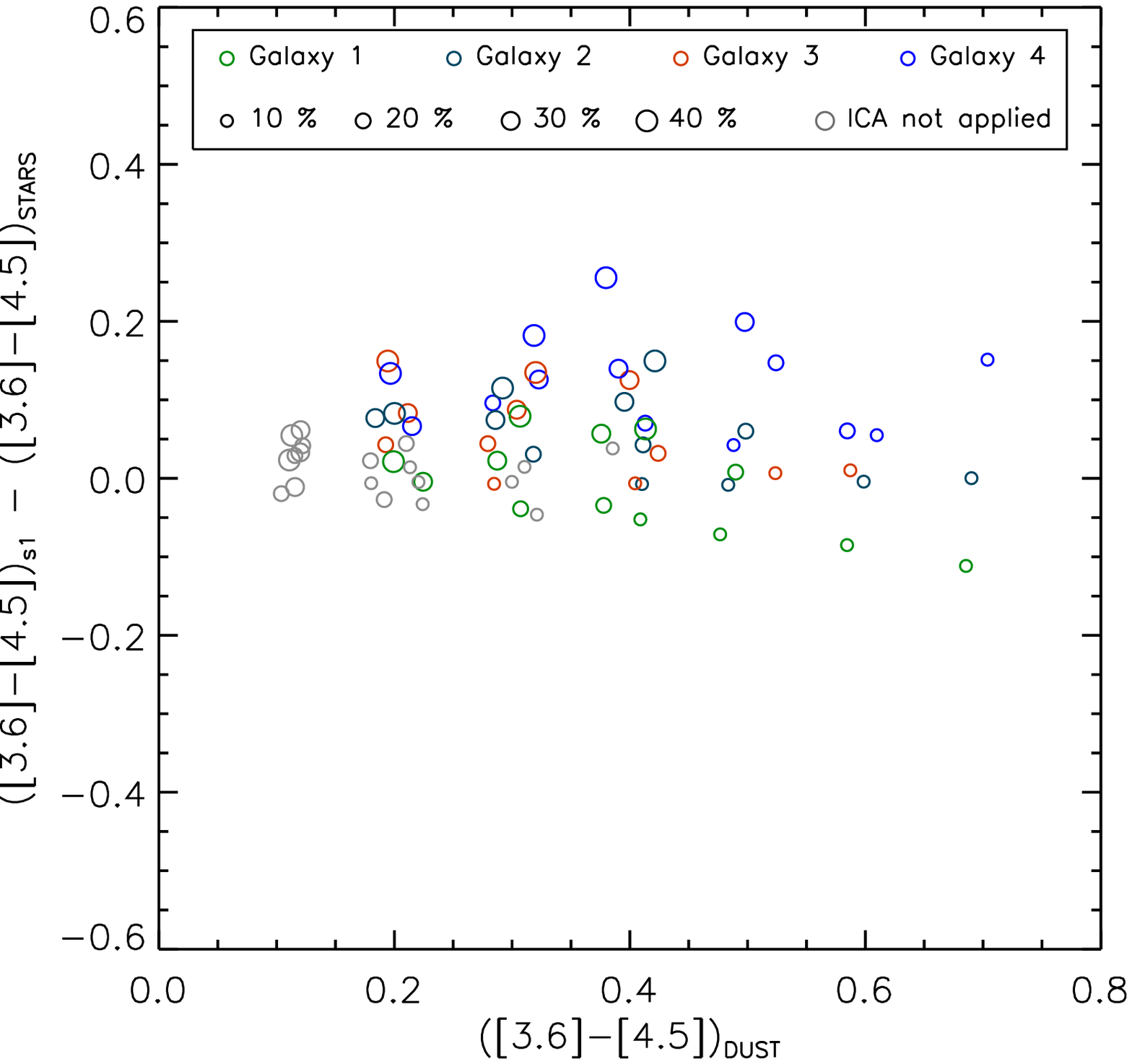}
\end{array}$
 \caption{Difference between the $s1$ (left) and $s2$ (right) colors identified by ICA and the original color of stars and dust in the models as a function of dust input color. Models where ICA would not have been applied due to original blue colors are shown in grey.}
\label{fig:2compA}
\end{center}
\end{figure}

The difference between the stellar and dust colors imposed on each model and the $s1$ and $s2$ colors found by ICA are shown in Fig.~\ref{fig:2compA}, as a function of the dust color of the model. To improve the visibility of the plotted data points, we introduce a small artificial random scatter around each discrete dust color value (a random dithering within $\pm 0.05$). The symbol size of each point is proportional to the dust fraction, while the colors denote the original galaxy on which the models are based (highlighting the effect of 2-dimensional dust distribution). Most striking is the clear trend for increased scatter in the ICA-determined dust colors (and stellar colors, although less so) when the `intrinsic' dust colors are redder. ICA underestimates the dust color more severely for low contamination fractions, and, conversely, it tends to overestimate the stellar color (but not as much) at the highest
contamination fractions. 

\begin{figure}[h!]
\begin{center}
\includegraphics[width=0.4\textwidth]{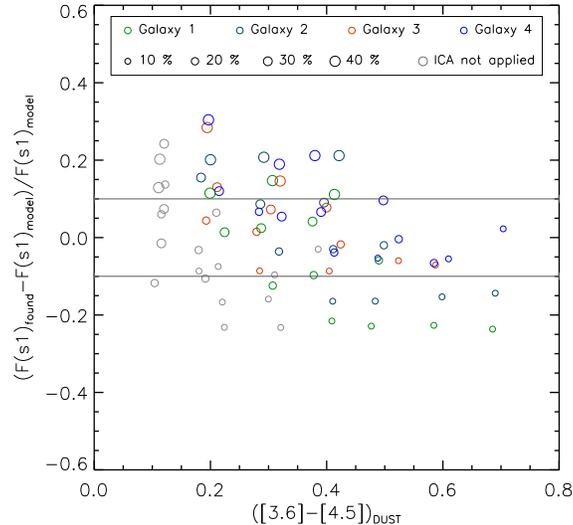}
\end{center}
\caption{Difference between the corrected stellar fluxes found by ICA and the stellar flux imposed on each of model, as a function of the imposed dust color. Symbol sizes are proportional to the dust contamination fraction, and different colors correspond to models based on different galaxies. Gray circles are those models that lead to original (weighted) mean colors that are negative, which means that we would not have applied ICA to them.}
\label{fig:2compB}
\end{figure}

However, as shown in Fig.~\ref{fig:2compB}, this large uncertainty in dust and stellar colors does not necessarily imply larger uncertainties in the estimation of fluxes. Plotted there is the difference between the stellar flux imposed on the model and the total flux in the $s1$ map retrieved by ICA for all possible combinations of dust colors and contamination fractions in our
model grids. Flux uncertainties in both components are sensitive to the relative stellar and dust fluxes (i.e. the level of dust contamination) but this tends to be balanced by an additional dependence on the actual difference between the stellar and dust colors.  As we are ultimately interested in the error in the corrected stellar fluxes found by ICA, Fig.~\ref{fig:2compB}
suggests that ICA can perform quite well. We note that many of the models exhibit original (weighted) mean [3.6]-[4.5] colors that are negative, so that we would not have applied ICA to them (see Sect.~\ref{subsec:icaneed}). Here we see that it is reasonable to exclude such galaxies from the analysis given that their low contamination fraction and/or very low dust colors (both of which can produce original colors $[3.6]-[4.5]<0$) lead to the highest uncertainties.

Fig.~\ref{fig:2compB} also shows that most models lead to final uncertainties in flux within 10\%, with some outliers covering a band up to 20\% of uncertainty. However, these are mainly associated with the highest contamination fractions in our models (40\%), and such extreme contamination from dust can be only very rarely expected (it is the upper limit for star-forming spirals, as argued in Sect.~\ref{fract_contam}). 

Additionally, these tests confirm that the ICA uncertainties determined through the bootstrap method are reliable measures of the errors in color. This bootstrap uncertainty is obtained by running ICA based on 48 different seeds (different starting points in $s1$ and $s2$ colors). According to our models, in 94.6\% of cases the difference in stellar color was well confined by the bootstrap uncertainty, confirming its meaningfulness.

\begin{figure}[h!]
\begin{center}$
\begin{array}{cc}
\includegraphics[width=0.4\textwidth]{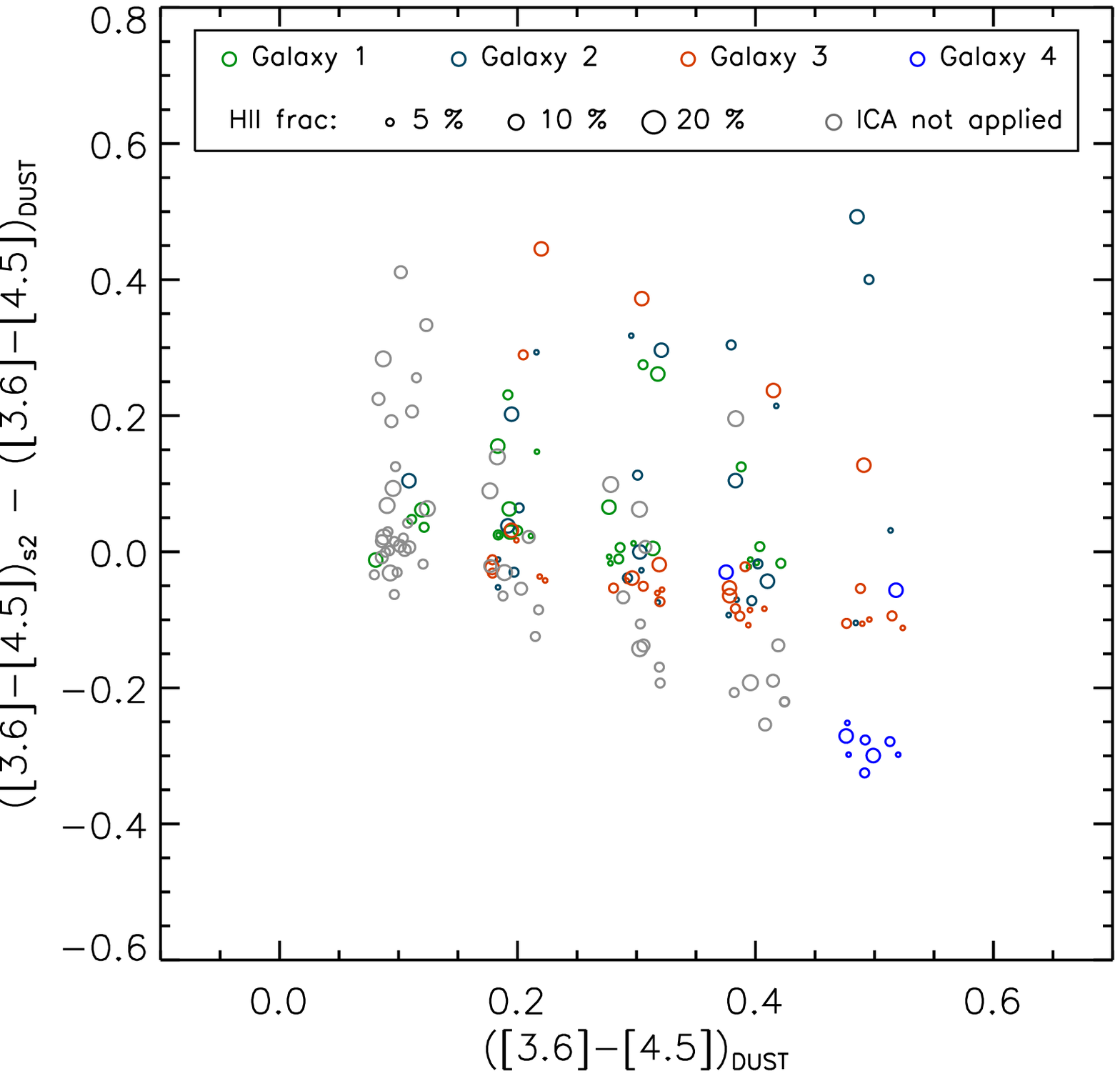} &
\includegraphics[width=0.4\textwidth]{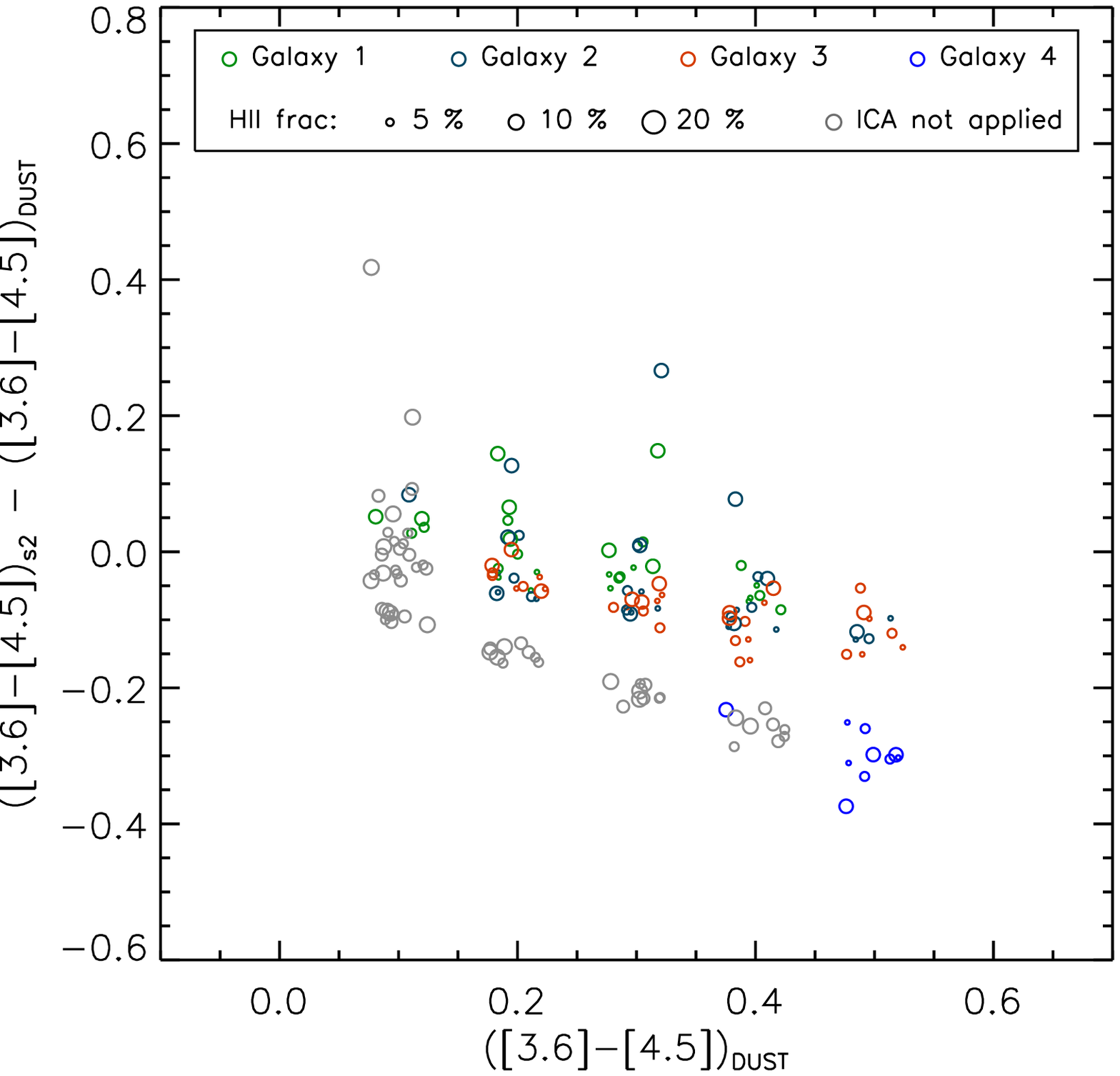} \\
\includegraphics[width=0.4\textwidth]{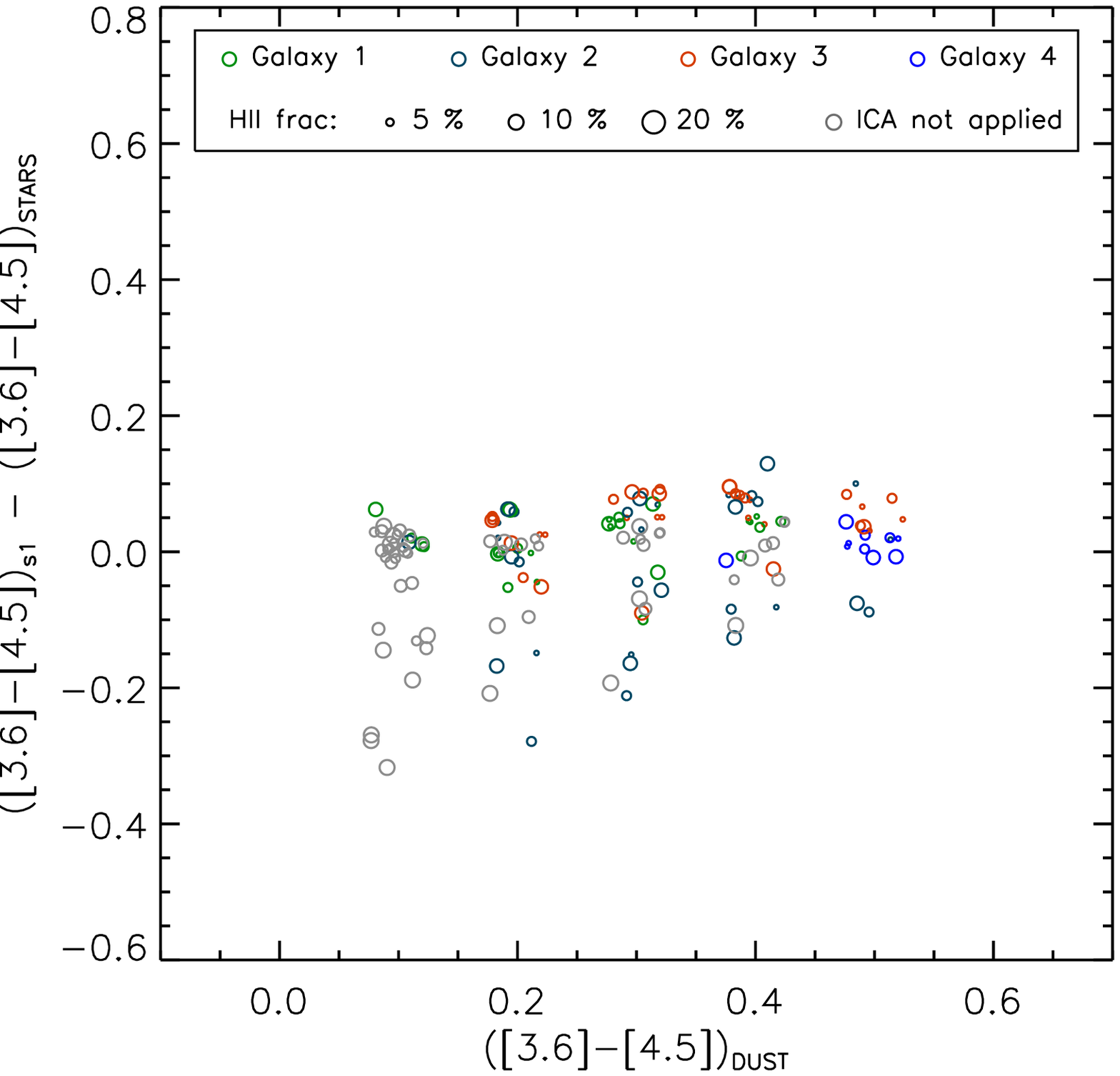} &
\includegraphics[width=0.4\textwidth]{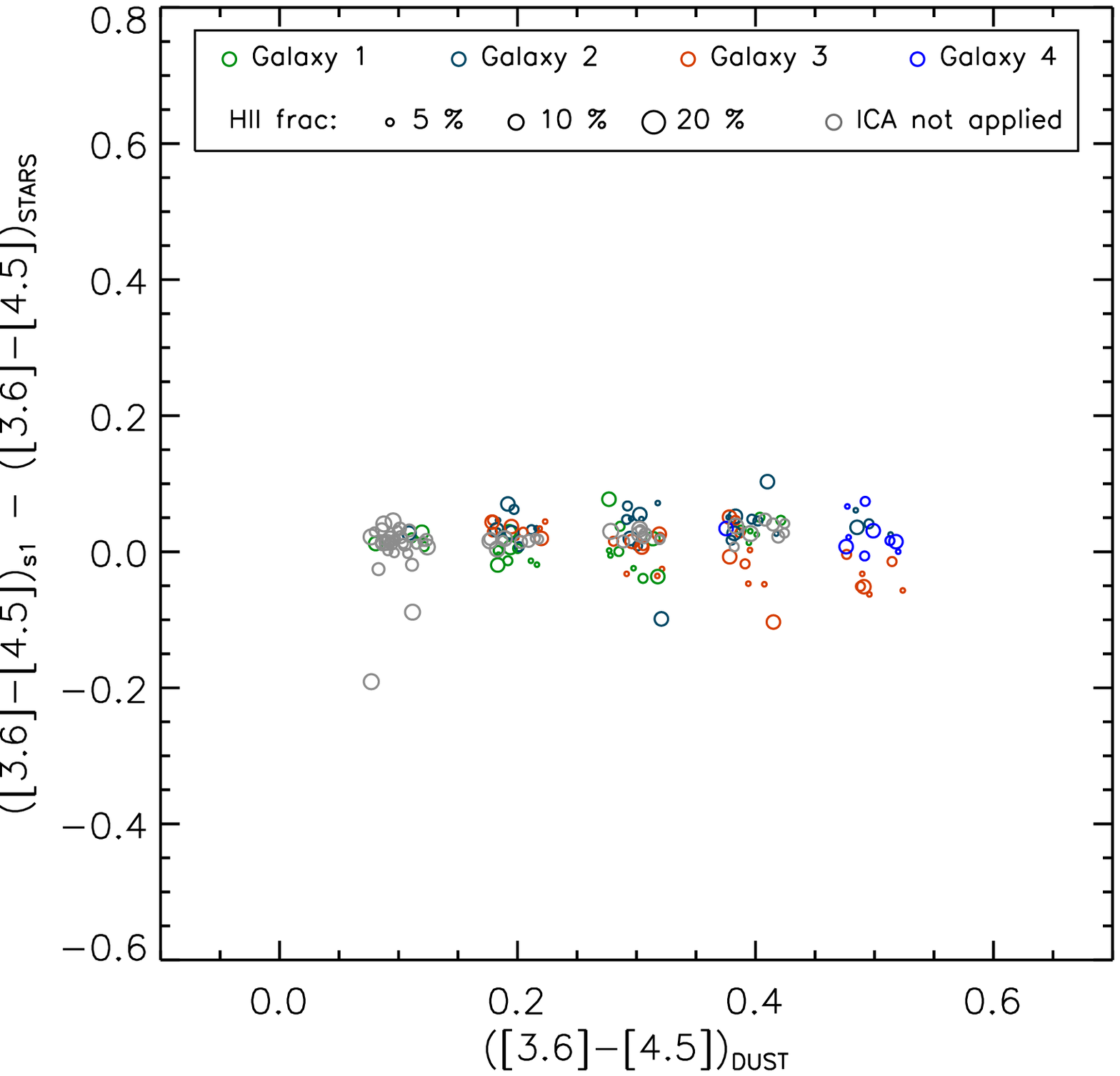}
\end{array}$
 \caption{Plots showing the effect of including a third component of redder colors in the models, and how our second iteration of ICA alleviates that problem. \textit{Left column}: first iteration, ICA$_1$; \textit{right column}: second iteration, ICA$_2$. \textit{Top}: difference between $s2$ color and the color imposed on the spatially extended dust, as a function of the color imposed on HII regions; \textit{bottom}: same difference as a function of the color imposed on the spatially extended dust.}
\end{center}
\label{fig:3comp}
\end{figure}

\section{Effect of second ICA iteration}

Given the possibility that more than one emitting dust component may be present (and, in particular, for HII regions in addition to the nominal diffuse dust component), we have implemented a second iteration of ICA (Sect.~\ref{ICA2}) aimed at removing the reddest emitting regions.  
Since ICA is sensitive to color outliers, a second iteration of ICA with the regions with more extreme colors masked allows us better retrieve the spatially dominant dust component, which is typically not as red. Our implementation of the second iteration is confirmed with the following tests, described below.

\subsection{Input models}

To test the impact of a third component on ICA solutions assuming two components (and to confirm the usefulness of our second ICA iteration), we have included an additional component with an intrinsically redder color representative of hot dust in HII regions, i.e. [3.6]-[4.5]=1. For the spatial flux distribution, we use truncated De Vaucouleurs profiles \citep{1948AnAp...11..247D}, and place them randomly in the regions of the galaxy model where the more extended dust emission is significant (higher than the mean dust flux). We vary the total number of regions (10, 20, 30) and their relative brightness (5\%, 10\%, 20\% of total dust flux). We also explore a range of colors, starting from the color of the spatially extended dust itself, up to a maximum color of 0.6.

\subsection{Findings}

When run through our pipeline, we can indeed see that our simulated HII regions get effectively masked, and the second iteration leads to a dust color closer to the nominal spatially extended `diffuse dust' in 94.7\% of the experiments (see Fig.~\ref{fig:3comp}). Moreover, the bootstrap uncertainty is effectively reduced after the second iteration (from $\sim 0.8$ to $\sim 0.2$~mag), suggesting that our measure of errors is sensitive to the problem posed by a third component of intrinsically different color, which is relieved by the iterative ICA approach that we have constructed.

\bibliography{/Users/querejeta/Documents/E1_SCIENCE_UTIL/mq.bib}{}
\end{document}